\documentstyle[referee]{l-aa}

\begin{document}

\thesaurus{0.7(19.11.1;19.15.1;19.34.1:19.43.1)}

\title{The age-mass relation for chromospherically active binaries}

\subtitle{II. Lithium depletion in dwarf components
\thanks{Tables 1, 2 and 3 are also available in electronic form at the
 CDS via anonymous ftp to cdsarc.u-strasbg.fr (130.79.128.5) or via
 http://cdsweb.u-strasbg.fr/Abstract.ht ml}
\thanks{Based on
 observations collected with the 2.2m telescope of the German-Spanish 
 Observatorio de Calar Alto (Almer\'\i a, Spain), and with the 2.56m Nordic 
 Optical Telescope in the Spanish Observatorio del Roque de Los Muchachos of 
 the Instituto de Astrof\'\i sica de Canarias (La Palma, Spain)}}

\author{D.~Barrado y Navascu\'es \inst{1,2} 
         \and M.J.~Fern\'andez-Figueroa \inst{3}
         \and R.J.~Garc\'{\i}a L\'opez \inst{4}
         \and E.~De~Castro \inst{3}
         \and M.~Cornide \inst{3}}

\offprints{D.~Barrado y Navascu\'es, dbarrado@cfa.harvard.edu}

\institute{MEC/Fulbright Fellow at the Smithsonian Astrophysical Observatory, 
	   60 Garden St, Cambridge, MA 02138, USA.
        \and
           Real Colegio Complutense at Harvard University,
	   Trowbridge St, Cambridge, MA 02138, USA.
        \and
	   Dpto. Astrof\'{\i}sica. Facultad de F\'{\i}sicas.
           Universidad Complutense. 
           E-28040 Madrid, Spain. 
        \and
           Instituto de Astrof\'\i sica de Canarias. E-38200 La Laguna, 
           Tenerife, Spain.}                     
  
\date{Received date, November 25th, 1996 ; MArch 1997  }

\maketitle

\begin{abstract} 

We present an extensive study of lithium abundances in dwarf components of 
chromospherically active binary stars (CABS). Since most of these binaries have 
known radii, masses and ages, this kind of data is especially useful for 
comparisons with theoretical models which try to explain the Li depletion 
phenomenon. We show that a significant part
 of these stars have clear Li overabundances with 
respect to the typical values for stars of the same mass and evolutionary stage. 
These excesses are evident when comparing our sample of CABS with binary and 
single stars belonging to open clusters of different ages, namely Pleiades, 
Hyades, NGC752, M67 and NGC188, which have ages ranging from 7$\times$10$^7$ to 
10$^{10}$ yr. The Li excesses are more conspicuous for masses in the range 
0.75--0.95 M$_\odot$, indicating that the rate of Li depletion has been less 
pronounced in CABS than in single stars. This phenomenon is interpreted in the 
context of transport of angular momentum from the orbit to the stellar rotation 
due to tidal effects. This angular momentum transfer would avoid the radial 
differential rotation and the associated turbulent mixing of material in the 
stellar interior. Other explanations, however, can not be ruled out. This is the 
case of transport of material induced by internal gravity waves, which could be 
inhibited due to the presence of strong magnetic fields associated with the 
effective dynamo in CABS. The confirmed existence of a relation between Li 
abundances and the fluxes in Ca\,{\sc ii} H\&K lines can also be accommodated 
within both scenarios.


\end{abstract}

\section{Introduction}

The  stellar 
group known as chromospherically active binary stars (CABS) is 
characterized by the presence of cool components which show emissions in 
Ca\,{\sc ii} H\&K, H$\alpha$, Mg\,{\sc ii} h\&k, C\,{\sc ii}, C\,{\sc iv} and 
other spectral lines. They have also other special characteristics, such as 
short orbital periods and high rotational velocities (Strassmeier et al. 1993), 
and include RS CVn binaries, which contain at least an evolved component, and BY 
Dra stars, which are still in the main sequence (MS). This last group comprises 
young single stars with rapid rotation and close binary stars. 

This paper belongs to an ongoing project designed to study the evolution and 
properties of CABS. In preliminary works (Fern\'andez--Figueroa et al. 1993; 
Barrado et al. 1993) we analyzed the Li abundances of a small sample, and in 
Paper I of this series (Barrado et al. 1994) we described the evolutionary 
status of this kind of binaries, and studied a subsample of CABS having accurate 
mass determinations (e.g. the eclipsing binaries). Using the position of each 
component on a radii--T$_{\rm eff}$ plane, we classified a group of 50 CABS in 3 
different types: MS stars having masses $<$1.7 M$_\odot$, evolved stars with 
masses around 1.4 M$_\odot$, and giants with masses in the range 2.5--5 
M$_\odot$. It is also possible to discriminate 3 different evolutionary stages 
in the second group: subgiants evolving off the MS, subgiants at the bottom of 
the red giant branch (RGB), and giants ascending the RGB above the previous 
subgroup. Using evolutionary tracks to compute the ages of these stars, we found 
that the ages depend essentially on the stellar mass, and 
that the actual values are quite close to the corresponding ages for the 
terminal age main sequence (TAMS). We interpreted this relation in the context 
of the evolution of the internal structure and rotation,
 as an effect of the increase of the stellar radius as the component 
evolve off 
the MS, and the decrease of the rotation period due to tidal effects.
On the other hand, since the 
components of CABS, (especially close binaries) do not spin down by magnetic 
breaking due to the transfer of angular momentum from the orbit to the stellar 
rotation, these stars have the needed ingredients to show an enhanced dynamo 
effect. 

We study here a subsample of CABS with dwarf components. From now on, we will 
use the word {\it dwarf} to denote stars of luminosity classes V and IV which 
are evolving off the MS, but they have not reached the RGB. We take now into 
account that the abundances of light elements can change in the stellar interior 
as a result of different nuclear reactions. In particular, lithium is destroyed 
inside main sequence stars by the reaction $^7$Li(p,$\alpha$)$^4$He when the 
temperature is higher than 2.6$\times$10$^6$ K. Late-type stars have a 
convective envelope which deepens with decreasing stellar mass, facilitating so 
the transport of material from the bottom of this well mixed zone to the Li 
burning layer. Because of this, Li abundances depend on mass and age for MS 
stars, dependence very well established by numerous studies made in open 
clusters and field stars (see, for a comprehensive review, Balachandran 1994). 
For the masses considered in this study, the actual internal stellar structure 
models do not predict temperatures high enough to burn Li at the 
bottom of the convection zone, and there should be additional mechanisms, beside 
pure convection, responsible of the transport of material below the convective 
envelope. The aim of this work is to help to discriminate between proposed 
theoretical mixing mechanisms. We deal here with the Li abundances of those 
components of CABS classified as MS stars and stars evolving off the MS, and we 
will present the study of Li abundances in evolved components of CABS in a 
forthcoming paper (Barrado y Navascu\'es et al. 1997b, Paper III).

Following Paper I, we start by studying the evolutionary status of CABS. In this 
case we have used all of the data contained in the Strassmeier et al.'s (1993) 
catalog.
We have based our discussion in the photometry of each component to 
classify the 
sample. 

We describe these details and present the 
selected sample of dwarfs in Section 2. The observations and data reduction 
process are presented in Section 3. Section 4 is devoted to the measurement of 
Li equivalent widths (EW) and the derivation of Li abundances. Section 5 
contains a comparison between CABS and single and binary stars belonging to 
different open clusters. The relations to other stellar parameters are
 shown in Section 6, and, finally, the summary and conclusions are 
presented in Section 7.

\section{Dwarf components of CABS}

\subsection{The evolutionary status}

In Paper I we studied the evolutionary status of CABS using a sample of binaries 
which have accurate values of stellar masses and radii. We concluded that CABS 
are in a very specific evolutionary stage, since we found a relation between 
their ages and the masses of the primary stars. 

Our goal here is to study in detail the status of the whole sample of CABS and, 
in particular, to define the subsample of dwarfs components. Since an important 
part of the systems included in the catalog by Strassmeier et al. (1993) has no 
eclipses (then the radii and masses are not well known or are not known at all), 
the study about the evolutionary status must be conducted in a different way 
than in Paper I. We have used here the absolute visual magnitudes (these systems 
are quite near to the Sun, and most of them have measured trigonometric 
parallaxes) versus the (B--V) colors to produce Figure 1. We have overplotted 
the evolutionary tracks by Schaller et al. (1992) and Schaerer et al. (1993a) 
for metallicities Z=0.020 (solid lines) and Z=0.008 (dashed lines), 
respectively. Dwarf, subgiant and giant primaries (namely, the brightest star in 
the binary system) are shown as solid pentagons, squares and triangles, whereas 
dwarf, subgiant and giant secondaries are shown as open pentagons, squares and 
triangles, respectively. 

In order to be able to locate most of the components of CABS in the 
color--magnitude diagram of Figure 1, we proceeded in different ways: for an 
important part of the CABS there are available values of (B-V) for each 
component, which were measured during eclipses. In other cases, the components 
have equal mass or spectral type. Then, the color should be the same and the 
magnitudes follow the relation $V_{\rm primary}$=$V_{\rm secondary}$=$V_{\rm 
observed}$+0.753$^{\rm m}$. Under certain conditions (when the binary is close 
to the ZAMS) it is possible to perform a deconvolution process of the 
photometry, as described by Barrado y Navascu\'es \& Stauffer (1996), to compute 
individual colors and magnitudes. In a few cases, absolute magnitudes were 
assigned using the spectral type. Details of this process can be found in the 
footnotes to Table 1.

As it can be seen in Figure 1, the spectral classification does not provide 
clear boundaries to separate evolutionary stages of CABS. Moreover, the 
distinction between hot and cool components, although useful to identify the 
stars inside the system, is quite confusing when the evolutionary status is 
studied. In fact, depending on the masses of both components, the cool star can 
be the more massive one (and then the most evolved one in the case of a 
K~III$+$F~V binary) or the vice versa (the case of a K~V$+$M~V binary). 
Therefore, we will use the terms primary and secondary to refer the more and 
less massive star, respectively. When the masses are not known, the primary is 
defined as the most evolved star.

According to Figure 1, it is possible to classify CABS in 5 different groups:

\begin{itemize} 

\item  Stars inside the MS.

\item  Stars evolving off the MS or crossing the giant gap (MS-off).

\item  Stars at the bottom of the red giant branch. 

\item Those ones in the top part of the RGB.

\item Components in the horizontal branch (HB).

\end{itemize} 

\noindent This classification attends to differences in the physical structure 
of the stars and the mechanisms they use to produce energy. Then, the CABS in 
the second group (MS-off) have finished the hydrogen combustion in the center of 
the core and are expanding or developing their convective envelopes. The 
difference between the third and fourth groups (B-RGB and T-RGB) arises from the 
fact that B-RGB stars are burning helium, whereas T-RGB stars  could burn also 
other elements. Note that this classification is in excellent agreement with 
that one described in Paper I based on the positions on the Radii--T$_{\rm eff}$ 
plane.

The age of each component of the CABS was computed using again the theoretical 
evolutionary tracks by Schaller et al. (1993) and  Schaerer  et al. (1993a,b). 
The method for estimating ages consists of an interpolation with radii and then, 
with masses, because both stellar parameters are well known for an important 
part  of  stars in our subsample of CABS containing dwarf components.  As shown 
in Paper I, there is a clear correlation between mass and age for CABS. We have 
used this characteristic to assign the ages of those CABS of our sample which 
have known values of their masses but their radii have not been measured. The 
actual expression is:

\begin{equation}
{\rm Log~Age = (9.90\pm0.20) - (2.84\pm0.06)~Log~(Mass/M_\odot)}.
\end{equation}

\noindent An illustration of this relation can be seen in Figure 3 of Paper I.

\subsection{Program stars}

The basic data --namely, photometry, stellar masses, radii, distances, orbital 
and rotational periods, etc-- were selected from the Catalog of CABS 
(Strassmeier et al. 1993 and references therein). Table 1 shows the systems 
which have been studied in this paper: column 1 lists the name of the system, 
column 2 the HD number, columns 3, 4, 5 and 6 
{\it our} assignation of the (U--B), 
(B--V), (V--R)$_{\rm J}$ and (V--I)$_{\rm J}$ colors, respectively, for each 
component. Column 7 lists the distance whereas the 8th, and 9th columns contain 
the apparent visual magnitude and the absolute magnitude computed with the 
previous 2 columns. In this last case, we also took into account other 
information in order to obtain the most accurate estimation of the magnitude of 
each component in the binary system (see notes in Table 1). Finally, column 10 
shows our evolutionary classification (MS or MS-off star), and column 11 
provides the spectral type. The last column contains notes about the method used 
to compute the absolute visual magnitude and the photometric indices for 
each component. In total, our sample has 41 systems, which contain 62 dwarf 
components. As far as we know, our subsample does not have any bias with respect 
to the whole sample of CABS.

\section{Observations and data reduction}

The spectra of the stars analyzed here were collected during three different 
observational runs. The two first ones were carried out at the coud\'e focus of 
the 2.2m telescope of the Calar Alto Spanish--German Observatory (CAHA) in June 
1993 and May 1994, respectively. The last run was carried out at the Nordic 
Optical Telescope (NOT, 2.56m) of the Spanish Roque de Los Muchachos Observatory 
(La Palma) in September 1994.

For the first observational campaign we used the Boller and Chivens spectrograph 
equipped with a FRD CCD (1024$\times$1024 pixels, 
13$\mu$m/pixel) at the F/3 camera and a GEC CCD 
(1155$\times$768 pixels, 22.5 $\mu$m/pixel) at the F/12 camera.
 A  slit of 1 arcsec was 
used, and the spectral resolution achieved (measured from the FWHM of Th--Ar 
emission lines spectra) was R$\sim$30$\,$000 in both cases.  For the second 
campaign we selected the RCA CCD at the F/3 focus, (1024$\times$1024, 15 
$\mu$m/pixel, R$\sim$30$\,000$) and the 
TEK CCD at the F/12 focus (1024$\times$1024, 
24 $\mu$m/pixel, R$\sim$50$\,000$). Finally, the observations at the NOT were 
performed using the IACUB echelle spectrograph (McKeith et al. 1993) at the 
Cassegrain focus of the telescope, with a Thomson $1024\times 1024$ CCD 
detector. A final resolution of about 40$\,$000 was obtained. 

Exposure times were estimated in order obtain signal--to--noise (S/N) ratios in 
the range 100--160. Since the visual magnitudes of our objects are very 
different (1$^{\rm m}$$\le$V$\le$11$^{\rm m}$), we took exposures from 5 s 
up to
3600 s. In order to avoid the presence of cosmic rays, several exposures were 
taken for the longest exposures until the required S/N ratio was achieved,
adding  them at the end of the reduction process.
  Although radial velocities are
important in some cases due to the fact that the orbital periods are short, 
the exposure times were not long enough to produce an important doppler 
broadening of the lines, since in those cases the rotation is itself quite
important and is by far the main source of the broadening.

We reduced the data using standard procedures (bias subtraction, flat--field 
correction, extraction of one-dimensional spectra and continuum normalization) 
and the MIDAS\footnote{Munich Image Data Analysis System is a program developed 
by European Southern Observatory, Garching} package. For the CAHA data we 
transformed the pixel scale into wavelength scale by comparison with 
Th--Ar lamp  spectra. 
For the NOT data,  we calibrated the spectra by comparison with photospheric lines of the solar  spectrum taken by observing the Moon with the 
same instrumentation. 
 The wavelength calibration is extremely important, 
since a considerable part of our binaries are double--lined systems and the 
theoretical positions of the Li features of both stars were computed 
previously to compare with the observed wavelengths.

\section{Lithium abundances}

\subsection{Equivalent widths}

The $^7$Li\,{\sc i} 6707.8 \AA{ } feature is in fact a spectroscopic doublet 
separated by 0.15 \AA{ } (6707.76 and 6707.91 \AA, Andersen et al. 1984). With 
the resolutions of our spectra, the doublet appears as a single line. Moreover, 
since most of our stars are rapid rotators, the Li features are blended with the 
Fe\,{\sc i} 6707.4 \AA{ } line. On the other hand, 28 out of 41 systems of our 
sample are SB2 systems (double--lined systems), and the Li\,{\sc i}$+$Fe\,{\sc } 
complex can also be blended with other spectral features arising from the other 
component.

In order to measure the Li equivalent width (EW), we have fitted Gaussian 
profiles to the spectra.  Phases were accounted for during this process.
 Using the published ephemerides (Strassmeier et al. 
1993 and references therein), we computed the positions of the lines
arising from each component and used this
information  when fitting and measuring the equivalent widths.

Different situations appeared when we were performing 
the fitting procedure:

\begin{itemize}

\item SB1 systems with vsin\,i$\la$23 kms$^{-1}$. The blends between 
Li\,{\sc i} and Fe\,{\sc i} lines were weak and we were able to separate the Fe 
contribution with the fitting procedure. We used the comparison between the 
measured EW(Fe\,{\sc i} 6707.4 \AA) with empirical values obtained from (B--V) 
color indices (Soderblom et al. 1990) as a quality control of the procedure. We 
also obtained new empirical relations between EW(Fe\,{\sc i} 6707.4 \AA) and the 
equivalent widths of other spectral lines such as Fe\,{\sc i} 6703.6 \AA\ \& 
Fe\,{\sc i} 6705.1 \AA{ } and different colors. 

\item SB1 systems with vsin\,i$>$23 kms$^{-1}$. We measured EW(Li+Fe) and 
eliminated the Fe contribution taking into account the EW of stars with the same 
(B--V) color, and the empirical fits described above between EW(Fe\,{\sc i} 
6707.4 \AA) and the EWs of other iron  lines.

\item SB2 systems where we measured in the spectrum  $\Delta$v$_{\rm r}$$>$23 
kms$^{-1}$ (where the difference of radial velocities between components is 
defined as $\Delta$v$_{\rm r}$ = $|$v$_{\rm r}^{\rm primary}$ -- v$_{\rm r}^{\rm 
secondary}|$) and no blend with other close lines from the other component. The 
Li features from both components appeared separated. We proceeded like in the 
first case.

\item SB2 systems with multiple blends and narrow lines. We fitted as many 
Gaussian curves as necessary to obtain the EW(Li).

\item SB2 systems with multiple blends and wide lines. We measured the whole EW 
and tried to take out the different contributions with empirical relations 
between lines and colors.

\end{itemize} 

\noindent Some examples of these fitting procedures can be found in Figure 2.
On the other hand, the continuum location is very important. There are 
uncertainties due to blends with other lines and the noise of the spectra. As a 
rule, we located the continuum in the middle of the noise distribution for each 
wavelength.

Since most of our spectra have a secondary component bright enough to contribute 
to the continuum of the primary star, a correction of the measured EW is 
necessary. The continuum correction factors --CCF-- were calculated using the 
equation: 

\begin{equation}
{\rm CCF = \cases{(1+\alpha) ,&for hot components\cr
              (1+\alpha)/\alpha ,&for cool components\cr}},
\end{equation}

\noindent where $\alpha$ is defined (Boesgaard \& Tripicco 1986b) as:

\begin{equation}
{\rm \alpha = \frac{e^{(2.146\times10^4/T_{\rm eff}^{\rm
hot})}-1}{e^{(2.146\times10^4/T_{\rm eff}^{\rm cool})}-1}\times
{R_{\rm cool} \overwithdelims () R_{\rm hot}}^2\qquad,}
\end{equation}

\noindent or, in the cases the radii ($R_{\rm hot}$, $R_{\rm cool}$) were not 
available, we used the relation between the fluxes derived from the apparent 
magnitude of the primary and secondary ($V_{\rm p}$, $V_{\rm s}$), defined as:

\begin{equation}
{\rm  \alpha = 10^{\rm -(V_{s}-V_{p})/2.5}.}
\end{equation}

\noindent The differences between both methods are irrelevant. 
Then, the final EWs were obtained using:

\begin{equation}
{\rm EW^{\rm corrected} = EW^{\rm measured}\times CCF.}
\end{equation}

\noindent The results are shown in Table 2. In columns 1 and 2 we list the name 
of the star and its HD number, while columns 3 and 4 contain the parameters 
--effective temperature and radius-- needed to compute the continuum correction 
factors (CCF), listed in column 8. The method to compute CCF is indicated in 
column 9. Columns 5, 6 and 7 provide the measured equivalent widths of Li\,{\sc 
i} 6707.8 \AA{ } plus Fe\,{\sc i} 6707.4 \AA, and only the iron line or the 
lithium doublet, respectively. In these two last cases, the values are not 
corrected by the effect of the continuum due to the presence of a companion 
star.

\subsection{Effective temperatures}

We have used different photometric temperature scales to compute the final 
effective temperatures. Since most of our stars have available (B--V) colors, 
this index has been our
 basic temperature indicator. We estimated these temperatures  with 
the scale provided by Thorburn et al. (1993). However, for the coolest 
stars in 
our sample, we selected  scales based on (V--I)$_{\rm J}$ (Cayrel at al. 1985) 
and (R--I)$_{\rm J}$  (Carney 1983), and the final  T$_{\rm eff}$ was computed as 
the average of these values.
 For some cases, photometry is not available 
and we have used only the spectral type to compute the effective temperature 
(Schmidt--Kaler 1982). The standar deviation of 
the differences between temperatures obtained
from color indices and from the spectral types, for those stars having both,
is 320 K. 

We verified that this procedure does not introduce any bias in the final
temperatures and abundances. In fact, we computed, when possible, the different
temperatures for the stars in our sample and compared them. Except for
stars having temperatures less than 4500 K, there are neither significant 
differences (the average value is 80 K) not any clear trend. The differences
in the lower range of temperatures are easy to understand since the Thorburn
et al. (1993) temperature scale was obtained using stars with higher
 temperatures.

\subsection{Lithium abundances and errors}

In order to 
estimate the Li abundances, we used the Pallavicini et al. (1987) 
curves of growth (COG), valid for 4500 $\le$ T$_{\rm eff}$ $\le$ 6500 K. In 
those cases where the effective temperature was lower, we used the COG computed 
by Soderblom et al. (1993). All these COG were computed under LTE conditions and a 
gravity value of Log g=4.5. 
Systematic differences between the results obtained with both sets of COG were 
avoided by modifying the last group following the technique described by Duncan 
\& Jones (1983). This consists in computing the differences between the curves 
obtained for a given temperature (in our case, 4500 K) and shifting a whole set 
with these values. Then, the positions of both curves at 4500 K would overlap. 
The maximum value of the differences, in the linear part of the curves at any 
temperature, is 0.1 dex.  

In addition to our data, we have included the available measurements present in 
the literature. Specifically, we have re--analyzed the data by Pallavicini et 
al. (1992), Fern\'andez--Figueroa et al. (1993),  and Randich et al. (1993) 
using similar procedures. The final corrected EW(Li) and abundances are listed 
in colums 4 and 5 of Table 3.  This table also contains information about the 
stellar masses, orbital and photometric periods, and ages (computed by 
interpolating radii and masses with isochrones or from the relation between ages 
and masses for CABS, see Eq. 1 in  Section 2).

In general, spectral synthesis provides more accurate abundances because it
allows to remove different effects, such as the blends with other lines at
similar wavelength. However, in the specific case of CABS, Randich et al.
(1993) have shown that there are no important differences between abundances
derived using COG and those obtained using synthetic spectra techniques. Other
studies show similar results  (eg. Barrado y Navascu\'es et al.  1997a). 
Moreover, under special circumstances it is not possible to compute a reliable
synthetic spectrum for some binaries because there is not accurate information.
Since the goal of this paper is to study the evolution of the Li abundance in a
large sample of dwarf components of CABS, it is reasonable to use COG to derive
Li abundances. On the other hand, as mentioned above, the use of COG allows us
to include in our  analysis the EW(Li) provided in the literature.

Abundances were derived assuming LTE conditions. There are 
two ways in which the abundances could be affected by stellar activity {\it per 
se}. On one hand, there could be significant NLTE effects due to the presence of 
the chromosphere. As shown by Houdebine \& Doyle (1995), the inward UV and 
optical flux can reheat the upper atmosphere and modify the continuum level. 
Therefore, the observed EW(Li) would be 
smaller than that of a non active star for 
the same abundance, metallicity and mass. This effect would be relevant when the 
H$\alpha$ line appears in emission and when the abundances are high, but 
none of our stars presents
 both characteristics simultaneously. Note that this effect is 
opposed to the overabundances we have found (see next subsection). Then, we 
consider that the excesses in the Li abundances are real, and the NLTE is a 
second order effect in the Li abundance determination for our stars. Moreover, 
Pavlenko et al. (1995) have shown that the differences between Li abundances 
calculated under LTE and NLTE conditions for 
stars with T$_{\rm eff}\le 5500$ K are significant only for very strong lines. 
On the other hand, stellar spots could modify the Li abundance determinations. 
There are two associated effects: first, the spots can change the photometric 
indices and, therefore, the inferred T$_{\rm eff}$ which will be used to compute 
the abundances. Second, the EW(Li) is larger in spots than in the quiet 
photosphere, as shown by solar observations (Giampapa 1984). However, these 
effects will not modify the abundances more than $\sim$0.30 dex altogether in 
the worst case (Barrado y Navascu\'es 1996).

In addition to the previously described phenomenon associated with the presence 
of stellar spots, there are other different sources of errors in the Li 
abundance determinations, such as the EW measurement process, the effective 
temperature scale adopted, the computed continuum correction factors, and the 
results provided by different curves of growth. We have estimated the errors in 
our sample as follows: for two typical stars with 6000 and 5000 K, we have 
corrected EW(Li)s of $\sim$50 and $\sim$20 m\AA, respectively. The error in 
these measurements is $\sim$20\%. These numbers are translated into errors in 
the abundances equivalent to 0.16 and 0.13 dex in each case. On the other hand, 
the most important source of error appears in the temperature determination. The 
average error in the adopted effective temperatures is 200 K, and it produces 
uncertainties of 0.22 and 0.24 dex, respectively, in the abundances for both 
temperatures. The uncertainty in the continuum correction factors can be 
estimated in 10\%, introducing additional errors of 0.06 and 0.05 dex. Finally, 
the errors of LTE curves of growth 
add another $\sim$0.13 dex (Pallavicini et al. 1987).
 The final errors can be computed using the 
expression:

\begin{equation}
{\rm \Delta\,Log~N(Li)_{\rm final} = \{\sum_i (\Delta\,
Log~N(Li)_i)^2\}^{1/2} },
\end{equation}

\noindent where $\Delta$\,Log~N(Li)$_{\rm i}$ are the
individual errors described above. For both temperatures,
the final value is 0.31 dex. If the possible presence of
spots is taken into account, it would lead to a maximum
error of 0.43 dex. As noted by Randich et al. (1993), although 
these errors can be very important for individual stars, they
cancel each other and do not affect to the conclusions reached 
when studying the whole sample of stars.

\subsection{Li abundance vs. effective temperature}

As shown by Cayrel et al. (1984), Balachandran et al. (1988), Garc\'\i a L\'opez 
et al. (1994), and several other works, Li abundances of stars belonging to open 
clusters are strongly related to the stellar mass or effective temperature. 
Figure 3a shows the Li abundances of our sample of dwarf components of CABS 
against T$_{\rm eff}$ --solid circles--. For comparison purposes, we have also 
included members of the Pleiades (70--120 Myr, Meynet 1993) and Hyades (600--800 
Myr; Mermilliod 1981, Gilroy 1989) open clusters. Abundances from Pleiades were 
selected from Butler et al. (1987),  Pilachowski \& Hobbs (1987), Boesgaard et 
al. (1988), Soderblom et al. (1993) and Garc\'\i a L\'opez 
et al. (1994), whereas those  from  Hyades stars were 
obtained from Boesgaard \& Tripicco (1986a), Rebolo \& Beckman (1988), Thorburn 
et al. (1993) and Barrado y Navascu\'es \& Stauffer (1996).
 Pleiades stars are plotted as open triangles, whereas Hyades stars are included 
as open circles. The average behavior of single members of this cluster is shown 
as a solid line.

A first inspection of this diagram shows that the Li abundance of CABS depends 
strongly on T$_{\rm eff}$, as happens in open cluster stars. The estimated 
abundances for CABS are in some cases similar to those present in Pleiades 
stars. It could be concluded then that these CABS  have Pleiades' age. Other 
CABS, having abundances smaller than the Hyades, would be older than 600--800 
Myr. Moreover, it would be possible to constrain the age of the cooler stars in 
our sample (4500 $\ge$ T$_{\rm eff}$ $\ge$ 3000) between 70 and 800 Myr. 
However, this interpretation does not seem to be correct, for two reasons: 
first, it is a well known fact that low mass CABS belong to the old disk 
population (Eker 1992), and, therefore, their ages are between 10$^9$ and 
5$\times$10$^9$  yr, depending on the velocity dispersion. Second, the analysis 
presented in Paper I shows that the ages of these systems are much older than 
the Hyades, at least for those stars having masses  larger than 0.8 M$_\odot$.

On the other hand, the scatter in the abundances of CABS is as large as that 
present in Pleiades, and much larger than that corresponding to Hyades stars. 
The scatter observed in late-type stars of the Pleiades has been associated with 
the effect that rotation can have in the Li depletion phenomenon (Garc\'\i a 
L\'opez et al. 1994). Young rapid rotators would inhibit the Li depletion during 
their premain sequence phase (PMS) due to the effect of rotation on the internal 
structure (Mart\'{\i}n \& Claret 1996). During the MS phase, the scatter could 
be due to the lack of transfer of angular momentum and material between the 
envelope and the core (Pinsonneault et al. 1992). It could also be very 
important the role of magnetic fields in the inhibition of transport of material 
to the stellar interior (Spruit 1987; Schatzman 1993).  Part of the scatter in 
the abundances observed in the Log N(Li)--T$_{\rm eff}$ plot for CABS is due to 
a dependence on the stellar mass. Since an important part of our sample are 
quite evolved objects, and in fact in several cases 
they are leaving the main sequence, it is possible to have several CABS of 
different masses at the same temperature, as it is shown in Figure 3b. We have 
also included in the figure the sample of nearby dwarfs studied by Favata et al. 
(1996), and the Li depletion isochrones by Chaboyer (1993; from top to bottom, 
for 50, 70, 300, 800 and 4500 Myr.). As noticed by Favata et al. (1996), the 
observed dependence of Li on the temperature is not real, it is introduced by 
the conversion on the EW upper limits into abundances. The comparison between 
these two samples of old dwarfs allows us to determine that CABS have undergone 
a process of inhibition of their Li depletion. 
This phenomenon was previously 
established by Randich and Pallavicini (1991) and 
Pallavicini et al. (1992), using a sample of CABS which contained 
essentially giant stars. Our study, which makes use of dwarfs and therefore 
allows direct comparisons with open clusters, clearly confirms and supports this 
result. A more extensive discussion will be carried out in Sections 5 and 6.

\section{Comparison with open clusters}

As noticed above, Figures 3a,b are affected by the evolution of some 
stellar parameters such as temperature when a star approaches or crosses over the 
TAMS. To avoid this and the possible consequences on the interpretation of the 
abundances, we have compared them against stellar masses (Figure 4). In the 
particular cases of CABS, these values have been computed directly in most of 
cases because there are  available accurate orbits for both components,
with uncertainties below $\sim$0.05 M$_\odot$. For a 
few binaries, we estimated the masses using the M$\times$sin$^3$i values 
(obtained from the radial velocities curves)  and the inclination of the orbit 
(from the rotational periods and vsin\,i), the position in 
color--magnitude diagrams or the radii--T$_{\rm eff}$ plane (see footnotes in 
Table 3). The masses of cluster stars were estimated from isochrone fitting, 
following Balachandran (1995).

Figure 4a compares the abundances against stellar masses for  our sample of CABS 
and Hyades binaries. Note that we have differentiated tidally locked binary 
systems (TLBS) from well separated systems (Barrado y Navascu\'es \& Stauffer 
1996). In fact, some of the Hyades TLBS have been catalogued as CABS by 
Strassmeier et al. (1993). Barrado y Navascu\'es \& Stauffer (1996) have studied 
the Li abundances of TLBS in the Hyades and concluded that they show clear 
overabundances when the orbital period is less than 8 days, in excellent 
agreement with the theoretical predictions by Zahn (1994). In our case, we have 
a larger sample of stars, which are older than those and have P$_{\rm orb}$ 
ranging from less than 1 day to more than one hundred.

Since CABS are, in general, older than Hyades stars, they have had more time to 
synchronize their orbital and rotational periods. Therefore, binaries with 
larger P$_{\rm orb}$ are TLBS. As in the Hyades, there is a relation between Li 
abundances and P$_{\rm orb}$, in the sense that, for the same mass, (e.g. 
$\sim$0.8 M$_\odot$), Li tends to be higher in the case of TLBS than in the case 
of non-TLBS. Due to the fact that  TLBS with P$_{\rm orb}>$8 d have achieved 
synchronization during their MS life time (Zahn 1994), the Li inhibition should 
have taken place, at least in part, during this stage of the evolution.    In  
Figure 4a, CABS 
show some differences with Hyades binaries: 
First, an important part of CABS have larger abundances than Hyades stars of similar
mass, despite the fact that the studied CABS --except 42 Cap-- have the same age 
than Hyades or they are older. Second, the Li abundance scatter 
is much larger, and in the particular case of $\sim$0.8 M$_\odot$ Li abundances 
change by a factor 16\,000. The abundance spread for a given mass in an old 
cluster is never larger that $\sim$1 dex. Since CABS of equal  masses have 
similar ages, the observed scatter should be related 
to other parameters. The presence of excesses in the Li abundances of at least 
some of our binaries at $\sim$0.8 M$_\odot$ can be related to the results found 
by Garc\'{\i}a L\'opez et al. (1994) for stars with similar masses in the 
Pleiades. They showed that the abundances of stars having this mass are quite 
sensitive to rotation, and that the fast rotators deplete less Li than the 
slow ones.

On the other hand, it is worth to notice the detection of Li in very low mass 
stars, in particular YY Gem and V1396 Cyg. However, these abundances should be 
taken with some caveats since at these low temperatures the spectrum of late 
type stars presents multiple lines, together TiO and other molecular bands which 
were not taken into account to measure  EW(Li+Fe). 

It is possible to conclude from Figure 4a that the Li depletion is also less 
pronounced in CABS than the Hyades for those stars located in the mid F dip 
(Boesgaard \& Tripicco 1986a), despite that the CABS are more evolved and  
should have started to mix the external material with material poor in Li from 
inside, in the so called dilution process (Iben 1965).

Figures 4b,c,d present a comparison between the abundances of  CABS with members 
of the NGC752, M67 and NGC188 open clusters, which have ages corresponding to 
1.7--2.0$\times$10$^9$ (Hobbs \& Pilachowski 1986a; Meynet et al. 1993), 
3.0-6.0$\times$10$^9$ (VandenBerg 1985; Hobbs \& Pilachowski 1986b; Spite et al. 
1987; Garc\'{\i}a L\'opez et al. 1988; Montgomery et al. 1993; Meynet et al, 
1993; Balachandran 1995; Barrado y Navascu\'es et al. 1997c), and  
6.5--10.0$\times$10$^9$ yr (VandenBerg 1985; Meynet et al. 1993; Hobbs \& 
Pilachowski 1986b), respectively. 
In these figures, open cluster stars are whown as open symbols, CABS having
 similar ages to the clusters appear as solid squares, and other CABS are
 overplotted as solid circles. These figures show, as in Figure 3, that the 
Li abundance scatter could be attributed to an age spread. However,
 we  have shown in 
Paper I that dwarf CABS are near the TAMS, and therefore they are much older,
 in  most of cases, than these clusters. 

It is very interesting to perform a comparison using different mass intervals. 
Except the CABS belonging to the Hyades, which are outside of the validity of 
the mass-age relationship for CABS due to their low mass (the relation was 
computed for stars having masses larger than 0.8 M$_\odot$), all CABS classified 
as MS are older than NGC752 (see Figure 4b). Only some CABS located in Figure 1 
above the MS (MS-off) have ages comparable to this cluster. These stars have 
masses in the range of the mid F dip, and should have severely depleted their Li 
due to one of the mechanisms that work in that region (see Balachandran 1995 and 
references therein) and also should have started to dilute its abundance due to 
the mixing of material as the star develops a deep convective envelope.

On the other hand, the comparison of masses $\sim$1.4-1.2 M$_\odot$, 
corresponding to the cool side of the Li dip, in Figure 4c, between the CABS and 
M67 shows some clues about the Li depletion phenomenon when binaries start to 
evolve off the MS. In this figure, binaries and single stars of M67 are 
represented as open triangles and squares, respectively.
  As it can be seen, there is not a 
clear correlation between binarity and Li abundances in M67, despite the fact 
that in average they have higher abundances than single stars. A more detailed 
study about the Li abundances in binaries of M67 can be found in Barrado y 
Navascu\'es et al. (1997c). In this mass range, which contains few CABS, only 
one shows a clear overabundance. Therefore, it seems that Li  depletion is not 
affected in a large extent for this mass range. However, the Li excesses are 
more conspicuous in the case of CABS having smaller masses, since they are older 
than M67. They present the Li doublet in their spectra, so it is 
possible to derive abundances, when most of the M67 stars have only upper limits 
due to the lack of this feature in their spectrum. Finally, Figure 4d contains 
data of the oldest open cluster considered: NGC188. The conclusion is the same 
one as above, dwarf CABS have abundances larger than the values corresponding to 
their age. These results agree with the study by Spite et al. (1994). They found 
that close binaries belonging to the old disk population or the galactic halo 
have systematicaly larger abundances than binaries with larger orbital periods 
or single stars. Other excesses for the abundances of dwarf binaries have been 
found by Ryan \& Deliyannis (1995).

Another way to demonstrate the excesses in the Li abundances of CABS is shown in 
Figure 5. Here, the ages are plotted against the average Li abundances for some 
mass ranges.  Symbol sizes increase  with stellar mass (see key).
Error bars (0.17 dex for the age) are also shown.  We also 
have included the average abundances for different  ages, computed from cluster 
stars  in three mass ranges (0.85 M$_\odot$ $\le$ Mass $<$ 0.95 M$_\odot$, 0.95 
M$_\odot$ $\le$ Mass $<$ 1.05 M$_\odot$, 1.05 M$_\odot$ $\le$ Mass $<$ 1.15 
M$_\odot$). These average values were obtained using Li abundances derived in 
single stars which belong to open clusters, such as Pleiades, UMa Group, Coma, 
Hyades, NGC~752, M67 and NGC~188. We have carried out several fits of the type 
Log N(Li) = A + B~Log~Age, one for each mass interval.  Despite their age, CABS 
have high Li abundances. In some cases much higher than those ones 
characteristic of single stars at the same age. The difference becomes extreme 
when the comparison is made considering stars with M $<$ 0.95 M$_\odot$, 
corresponding to the two smaller circle sizes and the two 
bottom fits. Then, we conclude that
 the Li depletion rate is completely 
different, and slower, in CABS that in single stars.

\section{Li abundance, stellar activity and rotation}

One possible explanation for the high Li abundances present on the surfaces of 
CABS is the creation of Li via spallation reactions during very active episodes, 
in particular flares. In fact, Pallavicini et al. (1992) and Randich et al. 
(1993) tried to use this hypothesis to explain the overabundances that they 
found in their samples of CABS, essentially composed by giant stars. Several 
checks of this hypothesis have been carried out, since the creation of Li would 
imply a specific ratio between the isotopes $^6$Li and $^7$Li. Also, energetic 
criteria seem to reject the possibility of creation of Li at large scale. 
However, $^6$Li has been detected in HD84937 (Smith et al. 1993, Hobbs \& 
Thorburn 1994), an F-type metal-poor dwarf, and the calculations performed for 
Population II stars by Deliyannis \& Malaney (1995) indicate that it is possible 
to create Li in enough amount during the flares to substitute the depleted 
material, but, in principle, this mechanism is restricted to dwarfs and 
subgiants which have very shallow convective envelopes. The creation of 
Li during flares should produce a relation between the Li abundances and some 
activity indicators such as H$\alpha$, Ca\,{\sc ii} H\&K, the Ca\,{\sc ii} 
infrared triplet, Mg\,{\sc ii} h\&k, etc. We have tried to find out if there are 
such  relations for our sample of CABS. From the available data in the 
literature, we have found no obvious relation between the X-ray luminosity 
(Dempsey et al. 1993), or H$\alpha$ (Montes  et al. 1996). The lack of relations 
would indicate that there is no relevant Li creation during flares. However, 
there is a relation between Li abundances and the surface fluxes in Ca\,{\sc ii} 
H\&K, as provided by Fern\'andez--Figueroa et al. (1994) and Montes et al. 
(1995). This relation only shows up when specific ranges of masses are selected. 
In Figure 6 we have plotted the Li abundances against Ca\,{\sc ii} H\&K fluxes 
for stars having masses lower than 1.25 M$_\odot$ and masses in the range 3.0 
$\le$ M $\le$ 5.0 M$_\odot$. (We have included data corresponding to giant 
components of CABS in this figure, see Paper III.) 
An obvious interpretation of this phenomenon would be that the presence of 
stellar spots, due to the high activity rates, could modify the measured Li 
equivalent widths, since the Li\,{\sc i} 6707.8 \AA{ } feature is very sensitive 
to variations in the temperature (almost all Li is ionized in the considered 
range of temperature). Therefore, a stellar surface covered in a large 
proportion by spots would have a Li equivalent width larger than that one 
corresponding to the quiet photosphere, for a given value of the abundance. If 
this effect would not be corrected, the computed abundance would be larger than 
the real value.

Several works have tried to estimate the importance of this effect. Soderblom et 
al. (1993) performed different calculations to see if the observed spread in the 
Log N(Li)-T$_{\rm eff}$ plane for the Pleiades cluster could be attributed to 
this situation. They obtained that only an extreme situation (when the fraction 
of the surface covered by spots --filling factor-- were very high) could account 
for the spread. Moreover, Barrado y Navascu\'es (1996), using solar observations 
of the Li\,{\sc i} 6707.8, K\,{\sc i} 7699 and Na\,{\sc i}  5896 \AA{ } lines in 
different solar regions, computed a grid of simulations under different 
conditions (spots of different temperatures, presence of plages, different 
filling factors for the spots and the plages). He found that normal situations 
for CABS (filling factor up to 0.30) could only introduce maximum errors 
smaller than 0.30 dex in the Li abundances. A different strategy was adopted by 
Pallavicini et al. (1993). They studied carefully four active stars, trying to 
find correlations between the photometric variability 
and changes in the shape and the equivalent widths of the Li feature. They found 
no proves of changes in the Li related to surface inhomogeneities. 

All these evidences, together with the fact that we have not measured  
abundances larger than the so called ``cosmic abundance'', which corresponds to 
the abundance observed in young objects which have not undergone any depletion 
process (Log N(Li)=3.2, Mart\'{\i}n et al. 1994), allow us to conclude that,
with some caveats, 
there is no significant creation of Li, if any at all, on the atmosphere of 
CABS, and that the surface inhomogeneities do not affect essentially the 
computed abundances.

Our interpretation of the relation between Li abundances and Ca\,{\sc ii} H\&K 
fluxes goes in other direction: the synchronization between the rotational and 
the orbital periods. This peculiarity would affect the transport mechanism in 
the stellar interior, inhibiting the depletion of Li. Figure 4a shows that TLBS 
have, in average, larger abundances than other CABS without coupling between the 
orbital and rotational periods. The stars undergoing a magnetic breaking by 
stellar winds would transfer angular momentum from the convective envelope to 
the radiative core to avoid intense radial differential rotations, and this 
angular momentum transfer is accompanied by mixing of material in the stellar 
interior (Pinsonneault et al. 1989, 1990, 1992; Chaboyer et al. 1995). If 
magnetic breaking is inhibitted in TLBS, similar stars in different systems 
deplete Li in a different amount, because they have different orbital periods 
and the orbit works as a source of angular momentum for the spin. On the other 
hand, the Rossby number (the relation  between the 
rotational period and the turnover time of the convective cells, N$_{\rm R}$) is 
strongly related to different activity indicators (Noyes et al. 1984). 
Then, the relation between Ca\,{\sc ii} H\&K fluxes and Li abundances in CABS is 
a result of the inhibition of Li depletion caused by high rotation (due to tidal 
forces present in CABS), while increasing the stellar activity.

There is an alternative explanation for the relation between activity and Li 
abundance and for the high abundances found in CABS. This explanation relies in 
the mixing due to internal gravity waves generated by the pressure fluctuation 
of the convective cells at the bottom of convective envelope on the stellar 
interior (Garc\'{\i}a L\'opez \& Spruit 1991; Montalb\'an 1994). Montalb\'an 
(1994) shows how this mechanism produces a macroscopic diffusion, which 
transports Li (and other elements) toward the internal layers, where it is 
destroyed. On the other hand, strong magnetic fields in the bottom of the 
convective envelope, which are responsible of the stellar activity, can inhibit 
the generation of gravity waves (Schatzman 1993). In a more general way, the 
presence of a strong magnetic field could also inhibit turbulent mixing 
mechanisms between the base of the convection zone and the stellar interior 
(Spruit 1987). This would explain the link between activity and Li abundance and 
also the high abundances obtained in CABS. The actual data cannot 
discriminate which mechanism is actually working. Only a detailed modeling of 
the evolution of Li abundances and other stellar properties will help to find 
the answer.

\section{Summary and conclusions}

We have presented a study of Li abundances in dwarf components of 
chromospherically active binary stars. The accuracy in the measurements of 
masses and radii for an important part of the binaries studied here, suggests 
the use of these data to check different theories about the Li depletion 
phenomenon. 

Since CABS as a group contains binaries in very different evolutionary stages, 
we first classified the totality of binaries listed in the catalog compiled by 
Strassmeier et al. (1993). This
classification was based on the position of each component in a color--magnitude 
diagram. We selected stars located in the MS and those beyond the terminal age 
main sequence, but only if they have not reached the bottom of the red giant 
branch.

The study of the Li abundances in our sample shows that these stars present 
abundances higher than the normal values characteristic of stars of the same 
mass and age. This conclusion was reached by comparing our CABS, whose ages were 
derived by isochrone-fitting with masses and radii, with open cluster
stars. The comparison between stars in the range 1.2--1.4 M$_\odot$ --the cool 
side of the Li gap-- belonging to the M67 and the CABS, of similar ages, shows 
that Li depletion is not very sensitive to binarity in this mass range. However, 
notorious overabundances appear for CABS with 0.75 $\le$ Mass $\le$ 0.95 
M$_\odot$.

We have also confirmed the presence of a trend between the surface flux in 
Ca\,{\sc ii} H\&K and the Li abundance for CABS in specific mass ranges. We have 
interpreted this relation and the Li excesses in the context of the transport of 
angular momentum from the orbit to the spin, due to a tidal effect. This 
transport produces high rotational rates in these old stars (they do not spin 
down as single dwarfs do). Because they are, essentially, near the TAMS, they 
have deep convective envelopes and, hence, low Rossby numbers (a very enhanced 
activity). On the other hand, the interchange of angular momentum also avoids 
the radial differential rotation between the convective envelope and the 
radiative core. Therefore, the turbulent mixing which may appear due to this 
effect in single stars, leading to Li depletion, would be inhibited and Li would 
be preserved till certain extent. In short, active stars would have less 
differential rotation and more Li photospheric abundance than less active stars, 
explaining the relation between Li and Ca\,{\sc ii} H\&K 
emission. However, the data do not rule out other explanations such as the 
transport induced by internal gravity waves and its inhibition due to strong 
magnetic fields present at the bottom of the convective envelope.

\acknowledgements{
This research has made use of the Simbad database, operated at CDS, Strasbourg, 
France. We greatly appreciate the comments and suggestion on this paper
 by the referee, R. Gratton. 
DBN acknowledges the support by the Universidad Complutense with a grant and by 
the Real Colegio Complutense at Harvard University. This work has been partially 
supported by the Spanish Direcci\'on General de Investigaci\'on Cient\'{\i}fica 
y T\'ecnica (DGICYT) under projects  PB92-0434-c02-01 and  PB94-0203.}

\newpage

\begin{table*}
\caption[  ]{Photometry and evolutionary classification of the sample}
\begin{tabular}{lrccccrrrrll}
\hline
    Name         &    HD  &  (U-B)&(B-V) &(V-R)$_{\rm j}$&(R-I)$_{\rm j}$&D& V & Mv(d,V) & Lum &  Sp.type    &Comments\\
     (1)         &   (2)  &   (3)   &   (4)   &   (5)   &    (6)  &  (7)  &   (8)   &    (9)  &  (10)   &(11)&(12)   \\
\hline      
13 Cet A   p  &     3196   &  0.080  & 0.570  & 0.470  & 0.290  &    21.3  &   5.200  &   3.558  &   off & \{F7V/\}     & a,kc      \\  
13 Cet A   s  &     3196   &   --    &  --    &  --    &  --    &    21.3  &    --    &    --    &   MS  &    --        &           \\  
CF Tuc     s  &     5303   &  0.180  &  --    &  --    &  --    &    54.0  &    --    &   4.4    &   MS  &   G0V        & c,j       \\  
CF Tuc     p  &     5303   &   --    & 0.735  & 0.626  & 0.472  &    54.0  &    --    &   3.748  &   off &   K4IV       & a,kc      \\  
BQ Hyi     p  &    14643   &  0.380  & 0.850  & 0.696  & 0.483  &    77.0  &   8.190  &   3.758  &   off &    G1Vp      & a,kc      \\  
UX For     p  &    17084   &  0.210  & 0.730  & 0.640  & 0.458  &    --    &   7.990  &   5.3    &   MS  &    G5-8V     & a,kc,j    \\  
VY Ari     p  &    17433   &  0.630  & 0.960  & 0.610  & 0.554  &    21.0  &   6.900  &   5.289  &   off &   K3-4V-IV   & a,kc      \\  
V471 Tau   p  &     --     &  0.410  & 0.870  & 0.750  & 0.520  &    59.0  &   9.710  &   5.886  &   MS  &      K2V     & a,d,i     \\  
AG Dor     p  &    26354   &  0.630  & 0.950  & 0.794  & 0.544  &    --    &   8.670  &   6.15   &   MS  &    K1Vp      & a,kc,j    \\  
EI Eri     p  &    26337   &  0.140  & 0.670  & 0.610  & 0.390  &    75.0  &   6.960  &   2.585  &   off &    G5IV      & a         \\  
V818 Tau   p  &    27130   &  0.330  & 0.775  &  --    & 0.254  &    44.7  &   8.313  &   5.158  &   MS  &     G6V      & a,i       \\  
V818 Tau   s  &    27130   &   --    &  --    &  --    &  --    &    44.7  &    --    &   7.73   &   MS  &     K6V      & j         \\  
BD+14 690  p  &    27691   &  0.090  & 0.560  &  --    & 0.330  &    --    &   7.000  &   4.4    &   MS  &    G0V       & a,j       \\  
 vB 69     p  &    28291   &  0.335  & 0.746  &  --    &  --    &    --    &   8.590  &   5.9    &   MS  &    K0V       & a,j       \\  
V833 Tau   p  &   283750   &  0.950  & 1.085  & 0.947  & 0.738  &    16.7  &   8.160  &   7.046  &   MS  &    dK5e      & a,kc      \\  
V808 Tau   1  &   283882   &  0.920  & 1.060  &  --    & 0.420  &    42.0  &  10.423  &   7.307  &   MS  &     K3V      & b         \\  
V808 Tau   2  &   283882   &  0.920  & 1.060  &  --    & 0.420  &    42.0  &  10.423  &   7.307  &   MS  &     K3V      & b         \\  
BD+64 487  1  &    30957   &   --    &  --    &  --    &  --    &    --    &   9.353  &   6.4    &   MS  &     K2V      & j         \\  
BD+64 487  2  &    30957   &   --    &  --    &  --    &  --    &    --    &   9.353  &   6.4    &   MS  &     K2V      & j         \\  
V1198 Ori  p  &    31738   &  0.200  & 0.710  &  --    &  --    &    60.0  &   7.120  &   3.229  &   off &    G5IV      & a         \\  
YY Gem        &      --    &  1.040  & 1.350  & 1.340  &  1.06  &    13.7  &   9.823  &   9.139  &   MS  &    dM1e      & b         \\
HR2814     p  &    57853   &  0.050  & 0.590  &  --    &  --    &    26.0  &   6.600  &   4.525  &   MS  &     F9.5V    & a         \\  
54 Cam     p  &    65626   &  0.165  & 0.620  &  --    &  --    &    38.0  &   6.500  &   2.5    &   off &     F9IV     & a,d,k     \\  
54 Cam     s  &    65626   &   --    &  --    &  --    &  --    &    38.0  &    --    &   2.88   &   off &     G5IV     & k         \\  
TY Pyx     p  &    77137   &  0.260  & 0.720  & 0.550  &  --    &    55.0  &   6.835  &   3.9    &   off &      G5IV    & a,d,e,h   \\  
TY Pyx     s  &    77137   &   --    & 0.760  &  --    &  --    &    55.0  &    --    &   3.8    &   off &      G5IV    & h         \\  
$\xi$ UMa B p &    98230   &  0.220  & 0.590  &  --    &  --    &     7.9  &   4.870  &   5.382  &   MS  &     G5V      & a         \\  
DF UMa     p  &     --     &   --    & 0.890  &  --    &  --    &    22.7  &  10.140  &   8.389  &   MS  &    dM0e      & a,i       \\  
DF UMa     s  &     --     &   --    &  --    &  --    &  --    &    22.7  &    --    &   12.3   &   MS  &    [dM5]     & j         \\  
CD-38 7259 s  &   101309   &   --    &  --    &  --    &  --    &    62.0  &    --    &   5.1    &   MS  &    G5V       & j         \\  
IL Com     1  &   108102   & -0.010  & 0.510  & 0.400  & 0.300  &    86.0  &   8.833  &   4.161  &   MS  &     F8V      & b         \\  
IL Com     2  &   108102   & -0.010  & 0.510  & 0.400  & 0.300  &    86.0  &   8.833  &   4.161  &   MS  &     F8V      & b         \\  
RS CVn     s  &   114519   &  0.090  & 0.420  &  --    &  --    &   180.0  &    --    &   2.6    &   off &     F4IV     & c,h       \\  
MS Ser     p  &   143313   &  0.630  & 0.940  &  --    &  --    &    30.0  &    --    &   6.215  &   MS  &     K2V      & d,e,i     \\  
MS Ser     s  &   143313   &  1.150  & 1.230  &  --    &  --    &    30.0  &    --    &   7.725  &   MS  &     K6V      & j         \\  
$\sigma$$^2$ CrB s  &   146361   &  0.000  & 0.470  & 0.500  &  --    &    21.0  &   5.700  &   4.089  &   MS  &     F6V      & a         \\  
$\sigma$$^2$ CrB p  &   146361   &   --    &  --    &  --    &  --    &    21.0  &    --    &   4.4    &   MS  &     G0V      & j         \\  
WW Dra     s  &   150708   &  0.040  & 0.600  &  --    &  --    &   180.0  &    --    &   2.304  &   off &     G2IV     & d,m       \\  
V792 Her   s  &   155638   &   --    & 0.450  &  --    &  --    &   310.0  &    --    &   2.349  &   off &    F2IV      & m         \\  
 Z Her     p  &   163930   &   --    & 0.590  &  --    &  --    &   100.0  &   7.230  &   2.9    &   off &     F4V-IV   & g,h       \\  
 Z Her     s  &   163930   &   --    &  --    &  --    &  --    &   100.0  &    --    &   3.5    &   off &      K0IV    & h         \\  
V772 Her   Aa &   165590   &  0.130  & 0.590  &  --    &  --    &    41.7  &    --    &   4.4    &   MS  &     G0V      & d,e,h     \\  
V772 Her   B  &   165590   &   --    &  --    &  --    &  --    &    41.7  &    --    &   5.1    &   MS  &     G5V      & h         \\  
ADS 11060C p  &   165590   &  1.130  & 1.360  &  --    &  --    &    41.7  &  11.373  &   8.272  &   MS  &     K7V      & b         \\  
ADS 11060C s  &   165590   &  1.130  & 1.360  &  --    &  --    &    41.7  &  11.373  &   8.272  &   MS  &     K7V      & b         \\  
V815 Her   p  &   166181   &  0.130  & 0.720  & 0.540  &  --    &    31.0  &   7.660  &   5.222  &   MS  &      G5V     & a,i       \\  
V815 Her   s  &   166181   &   --    &  --    &  --    &  --    &    31.0  &    --    &   9.6    &   MS  &    [M1-2V]   & j         \\  
BY Dra     p  &   234677   &  1.000  & 1.221  & 1.100  & 0.780  &    15.6  &   8.070  &   7.4    &   MS  &     K4V      & a,d,h     \\  
BY Dra     s  &   234677   &   --    &  --    &  --    &  --    &    15.6  &    --    &   8.6    &   MS  &    K7.5V     & h         \\  
V478 Lyr   p  &   178450   &  0.210  & 0.740  & 0.650  & 0.430  &    26.0  &   7.680  &   5.605  &   MS  &     G8V      & a         \\  
HR 7578    1  &   188088   &  0.940  & 1.050  & 0.830  & 0.480  &    15.0  &   6.913  &   6.753  &   MS  &    K2-3V     & b         \\  
HR 7578    2  &   188088   &  0.940  & 1.050  & 0.830  & 0.480  &    15.0  &   6.913  &   6.753  &   MS  &    K2-3V     & b         \\  
V1396 Cyg  p  &    --      &  1.080  & 1.517  &  --    & 0.910  &    15.9  &  10.130  &   9.280  &   MS  &     M2V      & a,d,i     \\  
V1396 Cyg  s  &    --      &   --    &  --    &  --    &  --    &    15.9  &    --    &   11.3   &   MS  &     M4Ve     & j         \\  
\hline
\end{tabular}
\end{table*}

\setcounter{table}{0}
\begin{table*}
\caption[  ]{Continuation. Photometry and evolutionary classification of the sample}
\begin{tabular}{lrccccrrrrll}
\hline
    Name         &    HD  &  (U-B)&(B-V) &(V-R)$_{\rm j}$&(R-I)$_{\rm j}$&D& V & Mv(d,V) & Lum &  Sp.type    &Comments\\
     (1)         &   (2)  &   (3)   &   (4)   &   (5)   &    (6)  &  (7)  &   (8)   &    (9)  &  (10)   &(11)&(12)   \\
\hline      
ER Vul     p  &   200391   &   --    & 0.680  &  --    &  --    &    46.0  &   7.270  &   4.62   &   MS  &     G0V      & a,d,h     \\  
ER Vul     s  &   200391   &   --    &  --    &  --    &  --    &    46.0  &    --    &   4.74   &   MS  &     G5V      & h         \\  
42 Cap     p  &   206301   &  0.200  & 0.650  &  --    & 0.240  &    34.0  &   5.170  &   2.513  &   off &    G2IV      & a         \\  
KZ And     p  &   218738   &  0.550  & 0.893  &  --    &  --    &    --    &   8.733  &   6.4    &   MS  &    dK2       & b,j       \\  
KZ And     s  &   218738   &  0.550  & 0.893  &  --    &  --    &    --    &   8.733  &   6.4    &   MS  &    dK2       & b,j       \\  
SZ Psc     s  &   219113   &  0.000  & 0.440  &  --    &  --    &   125.0  &    --    &   3.3    &   MS  &    F8IV      & h         \\  
KT Peg     p  &   222317   &  0.160  & 0.670  &  --    &  --    &    22.5  &   7.040  &   5.394  &   MS  &    G5V       & a,l,m     \\  
KT Peg     s  &   222317   &   --    &  --    &  --    &  --    &    22.5  &    --    &   7.820  &   MS  &    K6V       & l,m       \\  
\hline
\end{tabular}
$\,$ \\
Spectral type between brackets are asumed from the photometry.
a.- The color indices include the contribution of the secondary, except when the color for the secondary is given.
b.- Two equal stars: (U-B)$_p$=(U-B)$_s$=(U-B)$_{obs}$, V$_p$=V$_s$=V$_{obs}$+0.753.
c.- The hot component dominates (U-B).
d.- There is also available M$_V^p$, M$_V^s$.
e.- It is posible to deconvolve the photometry in V.
f.- Combined M$_V$.
g.- The hot component dominates (B-V).
h.- M$_V$(hot/cool) were calculated independently during the eclipses.
i.- M$_V$(primary) were calculated by a deconvolution with the M$_V$(total) and M$_V$(secondary):
V$_{\rm p}$ = V$_{\rm s}$ - 2.5 Log [10$^{\rm -(V_{\rm total}-V_{s})/2.5}$ - 1].
j.- M$_V$ assumed from the spectral type.
k.- From the Strassmeier et al. catalogue.
l.- Distance from M$_V$, V.
m.- M$_V$ from R, T$_{\rm eff}$, CB.
\end{table*}

\setcounter{table}{1}
\begin{table*}
\caption[  ]{\footnotesize Equivalent widths and corrections for the observed stars (in m\AA).}
\begin{tabular}{lrccccccl}
\hline
 Name           &     HD & Teff &Radius    &EW(Li+Fe)& EW(Fe) &
EW(Li) & CCF(c/h)& Comments  \\
                &        & (K)&(R$_{\odot}$)& (m\AA)  & (m\AA)&
(m\AA) &         &    \\
     (1)         &    (2) &  (3) &    (4)  &    (5)  &   (6)  &  
(7)   &    (8)  &   (9)   \\
\hline                                                            
                                                       
  13 Cet A       &   3196 & 6023 &     --  &      89 &
4$^{calc}$&85$^{calc}$&      -- &        \\ 

  VY Ari         &  17433 & 4831 &     --  &      62
&23$^{calc}$&39$^{calc}$&      -- &        \\ 

  EI Eri         &  26337 & 5555 &     --  &      67 &     11 &   
 55 &      -- &        \\ 

  BD+64 487      &  30957 & 4900 &     --  &      54 &     16 &   
 38 & 2.000   &   3    \\ 

  BD+64 487      &  30957 & 4900 &     --  &    13.5 &      3 &   
 13 & 2.000   &   3    \\ 

  V1198 Ori      &  31738 & 5517 &     --  &      13 &      3 &   
 11 &      -- &        \\ 

  V1198 Ori      &  31738 &   -- &     --  &      -- &     -- &   
 -- &      -- &        \\ 

  YY Gem         &  --    & 3185 &     --  &      43 &
7$^{calc}$&36$^{calc}$& 2.000   &   3    \\ 

  YY Gem         &  --    & 3185 &     --  &      -- &     -- &   
 -- & 2.000   &   3    \\ 

$\xi$ UMa B       &  98230 & 5948 &     --  &      44 &      1 &   
 33 &      -- &        \\ 

  DF UMa         &  --    & 5117 &     --  &      68
&14$^{calc}$&52$^{calc}$& 1.027   &   3    \\ 

  DF UMa         &  --    & 3240 &     --  &      27
&16$^{calc}$&11$^{calc}$& 37.678  &   3    \\ 

  MS Ser         & 143313 & 4989 &     --  &      46
&40$^{calc}$& 7$^{calc}$& 1.249   &   3    \\

  MS Ser         & 143313 & 4275 &     --  &      -- &     -- &   
 -- & 5.018   &   3    \\ 

$\sigma$$^2$ CrB     & 146361 & 6444 &   1.22  &     30* &
4$^{calc}$&18$^{calc}$& 1.777   &   1    \\ 

$\sigma$$^2$ CrB     & 146361 & 6030 &   1.21  &     60* &
7$^{calc}$&30$^{calc}$& 2.287   &   1    \\ 

  WW Dra         & 150708 & 5910 &   2.12  &      87 &      4 &   
 34 & 2.158   &   1    \\ 

  WW Dra         & 150708 & 4580 &   3.9   &      21 &     13 &   
  7 & 1.863   &   1,4  \\    

  V772 Her       & 165590 & 5948 &   1.0   &      -- &     --
&105$^{calc}$&1.328   &   1    \\ 

  V772 Her       & 165590 & 5770 &   0.6   &      -- &     --
&83$^{calc}$& 4.114   &   1    \\ 

  ADS 11060C     & 165590 & 3979 &     --  &      43
&33$^{calc}$&10$^{calc}$& 2.000   &   3    \\ 

  ADS 11060C     & 165590 & 3979 &     --  &      -- &     -- &   
 -- & 2.000   &   3    \\ 

  V815 Her       & 166181 & 5496 &   0.97  &     202
&10$^{calc}$&192$^{calc}$&1.040   &   1    \\ 

  V815 Her       & 166181 & 3648 &   0.525 &      -- &     -- &   
 -- & 26.113  &   1    \\ 

  V1396 Cyg      &     -- & 3752 &   0.4   &      52 &     40 &   
 12 & 1.383   &   1    \\ 

  V1396 Cyg      &     -- & 3370 &   0.3426&      -- &     -- &   
 -- & 3.611   &   1    \\ 

  KZ And         & 218738 & 5117 &     --  &      29 &
7$^{calc}$&27$^{calc}$& 2.000   &   3    \\ 

  KZ And         & 218738 & 5117 &     --  &      34 &
7$^{calc}$&36$^{calc}$& 2.000   &   3    \\ 

  KT Peg         & 222317 & 5528 &   0.93  &      17 &     12 &   
  4 & 1.168   &   1    \\ 

  KT Peg         & 222317 & 4178 &   0.72  &      -- &     -- &   
 -- & 6.937   &   1    \\ 
\hline                           
\end{tabular}    
$\,$\\
* Line blended with others.
$^{1}$ CCF from Equation 2 and 3. 
$^{2}$ CCF from Equation 2 and 4.
$^{3}$ We deconvolved the magnitude of the primary from the total
magnitude and an estimation of the magnitude of the secondary by
using the spectral type.  
$^{4}$ Evolved components.
\end{table*}

\setcounter{table}{2}
\begin{table*}
\begin{flushleft}
\caption[ ]{\footnotesize Final abundances, where Log N(Li)=12$+$Log\{N(Li)/N(H)\},
 and other parameters for CABS}
\begin{tabular}{lrccrcllc}
\hline
  Name         &     HD &T$_{\rm eff}$&EW$_{\rm cor}$&N(Li)&  
Mass         & P$_{\rm orb}$   &P$_{\rm phot}$& Log Age        \\
                 &          &    (K)  &  (m\AA)   &           & 
(M$_\odot$)   &      (d)        &     (d)      &  (yr)     \\     
    (1)          &    (2)   &    (3)  &    (4)   &     (5)   &    
 (6)       &       (7)       &      (8)     &        (9)     \\
\hline
 13 Cet A p      &    3196  &   6023  &  85      &    
2.84&$\sim1.60^{\rm CM}$&2.08200 &$\sim2.08^{\rm Porb}$ &
9.221$^{\rm M}$\\  
 13 Cet A s      &    3196  &   5700  &$\le$1    & $\le$0.0  &   
--          & 2.08200 &$\sim2.08^{\rm Porb}$&$\sim9.22^{\rm P}$\\  
 CF Tuc   s      &    5303  &   6030  &  61.29   &     2.59  &   
1.057       &    2.79786      &    2.798     &      9.892     \\  
 CF Tuc   p      &    5303  &   5097  &  73.49   &     1.87  &   
1.205       &    2.79786      &    2.798     &      9.812   \\  
 BQ Hyi   p      &   14643  &   5140  &  26.64   &     1.27
&$\sim1.32^{\rm CM}$& 18.379        &    18.24 & $\sim9.454^{\rm
M}$\\  
 UX For   p      &   17084  &   5357  &  24.49   &     1.45  &   
--          &    0.95479      &    0.957     &      --       \\  
 VY Ari   p      &   17433  &   4831  &  39      &     1.10
&$\sim0.80^{\rm CM}$& 13.198        &    16.42 & $\sim10.059^{\rm
M}$\\  
 V471 Tau p      &    --    &   4752  &  218.7   &     2.72  &   
0.8         &    0.52119299   &    0.5197    &       Hyades   \\  
 AG Dor   p      &   26354  &   4964  &  16.46   &     0.82 
&$\le$0.80       &    2.562        &    2.533     &     --      \\  
 EI Eri   p      &   26337  &   5555  &  25.65   &     1.67  &   
--          &    1.947227     &    1.945     &      --       \\  
 vB22     p      &   27130  &   5542  &  67.3    &     2.28  &   
1.086       &5.609198 &$\sim5.61^{\rm Porb}$ &       Hyades   \\  
 vB22     s      &   27130  &   4366  &  --      &     --    &   
0.776       &5.609198 &$\sim5.61^{\rm Porb}$ &       Hyades  \\  
 vB40A    p      &   27691  &   6225  &  122.3   &     3.46  &   
--          &4.00000  &$\sim4.00^{\rm Porb}$ &       Hyades  \\  
 vB 69    p      &   28291  &   5539  &  11.8    &     1.28
&$\sim0.96^{\rm CM}$&  41.66        &    --        &       Hyades \\  
 V833 Tau p      &  283750  &   4540  &  16.2    &     0.23  &   
0.8         &    1.787797     &    1.797     &       Hyades   \\  
 vB117    p      &  283882  &   4650  &$\le$4.2  &$\le$-0.25
&$\sim0.79^{\rm Msin^3i}$& 11.9293 &    6.82      &       Hyades  \\  
 vB117    s      &  283882  &   4520  &$\le$2.6  &$\le$-0.65
&$\sim0.77^{\rm Msin^3i}$& 11.9293 &    6.82      &       Hyades  \\  
 BD+64 487 1     &   30957  &   4900  &  76      &     1.64 
&$\le$0.80       &    44.38        &    --        &      --    \\  
 BD+64 487 2     &   30957  &   4900  &  27      &     0.98 
&$\le$0.80       &    44.38        &    --        &      --    \\  
 V1198 Ori p     &   31738  &   5517  &  11      &     1.23 
&$\sim1.84^{\rm CM}$&    --        &    4.55   &$\sim9.059^{\rm
M}$\\  
 HR 2814   p     &   57853  &   5948  &  101.54  &     2.93  &    
--         &    122.169      &    --        &        --      \\  
 YY Gem          &   --     &   3185  &   65     &     0.11  &   
0.62        &    0.8142822    &    0.8143    &      --       \\
 54 Cam    p     &   65626  &   5829  &  65.70   &     2.47  &   
1.64        &    11.06803     &    10.163 &$\sim9.194^{\rm
M}$\\  
 54 Cam    s     &   65626  &   5460  &  116.57  &     2.71  &   
1.61        &    11.06803     &    10.163 &$\sim9.217^{\rm
M}$ \\  
 TY Pyx    p     &   77137  &   5438  &  4.53    &     0.75  &   
1.22        &    3.198548     &    3.32      &      9.730    \\  
 TY Pyx    s     &   77137  &   5390  &  7.05    &     0.89  &   
1.20        &    3.198548     &    3.32      &      9.796  \\  
$\xi$ UMa B  p     &   98230  &   5948  &  33      &     2.14  &    
--         &    3.9805  &$\sim3.98^{\rm Porb}$&      --      \\  
 DF UMa    p     &    --    &   5117  &  53      &     1.60  &   
0.56        &    1.033824&$\sim1.03^{\rm Porb}$&    --       \\  
 DF UMa    s     &    -     &   3240  &  --      &      --   &   
0.3         &    1.033824&$\sim1.00^{\rm Porb}$&    --       \\  
 CD-38 7259 s    &  101309  &   5770  &  41.54   &     2.13  &    
--         &    11.710       &    11.66     &        --      \\  
 IL Com    p     &  108102  &   6266  &  22      &     2.25  &   
0.85        &    0.9616       &    0.82      &      10.054   \\  
 IL Com    s     &  108102  &   6266  &  28      &     2.37  &   
0.82        &    0.9616       &    0.82      &      10.121   \\  
 RS CVn    s     &  114519  &   6700  &  --      & $\sim$3   &   
1.41        &    4.797851     &    4.7912    &      9.410    \\  
 MS Ser    p     &  143313  &   4989  &   9      &     0.57  &   
--          &    9.01490      &    9.60      &       --       \\  
 MS Ser    s     &  143313  &   4275  &  --      &     --    &   
--          &    9.01490      &    9.60      &       --      \\  
$\sigma$$^2$ CrB s    &  146361  &   6444  &  30      &     2.61  &   
1.12        &    1.139791     &    1.1687    &      9.611     \\  
$\sigma$$^2$ CrB p    &  146361  &   6030  &  69      &     2.67  &   
1.14        &    1.139791     &    1.1687    &      9.648    \\  
 WW Dra    s     &  150708  &   5910  &  48      &     2.37  &   
1.36        &    4.629617     &    4.63      &      9.575   \\  
 V792 Her  s     &  155638  &   6539  &  81      &     2.4   &   
1.41        &    27.5368      &    27.07     &      9.422   \\  
 Z Her     p     &  163930  &   5948  & --       &    --     &   
1.61        &    3.992801     &    3.962     &      9.012   \\  
 Z Her     s     &  163930  &   5000  & --       &    --     &   
1.31        &    3.992801     &    3.962     &      9.675  \\  
 V772 Her  Aa    &  165590  &   5948  &  158     &     3.19  &   
1.04        &    0.879504     &    0.878     &      9.465   \\  
 V772 Her  B     &  165590  &   5770  &  243     &     3.60  &   
0.88        &    7391.25      &    --        &      9.465   \\  
 ADS 11060C p    &  165590  &   3979  &  16      &    -0.43 
&$\sim0.70^{\rm Msin^3i}$ & 25.762 &    --        &     V772 Her  \\  
 ADS 11060C s    &  165590  &   3979  &  --      &     --   
&$\sim0.67^{\rm Msin^3i}$ & 25.762 &    --        &     V772 Her  \\  
 V815 Her  p     &  166181  &   5496  &  201     &     3.63 
&$\sim0.86^{\rm RT}$& 1.809837     &    1.819   &$\sim9.972^{\rm
M}$\\  
 V815 Her  s     &  166181  &   3648  &  --      &     --    &   
--          &    1.809837     &    1.819   &$\sim9.972^{\rm P}$ \\  
 BY Dra    p     &  234677  &   4300  &  7       &    -0.42 
&$\sim$0.55      &    5.975112     &    3.827     &      --       \\  
 BY Dra    s     &  234677  &   4021  &  8       &    -0.65  &   
0.44        &    5.975112     &    3.827     &      --        \\  
 V478 Lyr  p     &  178450  &   5395  &  59      &     1.97 
&$\sim0.83^{\rm RT}$& 2.130514     &    2.13    &$\sim10.014^{\rm
M}$\\  
 HR 7578   1     &  188088  &   4839  &  4.91    &     0.1  
&$\ge$0.80       &    46.815       &    16.5      & $\le$9.966 \\  
 HR 7578   2     &  188088  &   4839  &  4.91    &     0.1  
&$\ge$0.80       &    46.815       &    16.5      & $\le$9.966    \\  
 V1396 Cyg p     &   --     &   3752  &  17      &    -0.60  &   
0.42        &    3.276188 &$\sim3.28^{\rm Porb}$&     --     \\  
 V1396 Cyg s     &   --     &   3370  &  --      &     --    &   
0.27        &    3.276188 &$\sim3.28^{\rm Porb}$&     --     \\  
\hline
\end{tabular}
\end{flushleft}
\end{table*}

\setcounter{table}{2}
\begin{table*}
\begin{flushleft}
\caption[ ]{Continuation. Final abundances, where Log N(Li)=12$+$Log\{N(Li)/N(H)\},
 and other parameters for CABS}
\begin{tabular}{lrccrcllc}
\hline
  Name         &     HD &T$_{\rm eff}$&EW$_{\rm cor}$&N(Li)&  
Mass         & P$_{\rm orb}$   &P$_{\rm phot}$& Log Age        \\
                 &          &    (K)  &  (m\AA)   &           & 
(M$_\odot$)   &      (d)        &     (d)      &  (yr)     \\     
    (1)          &    (2)   &    (3)  &    (4)   &     (5)   &    
 (6)       &       (7)       &      (8)     &        (9)     \\
\hline
 ER Vul    p     &  200391  &   5602  &$\le$8    &$\le$1.15  &   
1.10        &    0.6980951    &    0.6942    &      9.569     \\  
 ER Vul    s     &  200391  &   5770  &$\le$9    &$\le$1.35  &   
1.05        &    0.6980951    &    0.6942    &      9.748     \\  
 42 Cap    p     &  206301  &   5710  &  42.07   &  2.09 
&$\sim2.58^{\rm CM}$& 13.1740 &$\sim13.17^{\rm
Porb}$&$\sim8.649^{\rm M}$\\  
 KZ And    p     &  218738  &   5117  &  58      &    
1.65&$\sim0.66^{\rm Msin^3i}$& 13.1740   &    3.03      &    --    \\  
 KZ And    s     &  218738  &   5117  &  62      &     1.68  &   
0.63        &    3.032867     &    3.03      &    --          \\  
 SZ Psc    s     &  219113  &   6500  & 75.50    &     3.31  &   
1.28        &    3.032867     &    3.955     &      9.496     \\  
 KT Peg    p     &  222317  &   5528  &  4       &     0.87  &   
0.93        &    6.201986     &    6.092     &      9.926    \\  
 KT Peg    s     &  222317  &   5528  &  --      &     --    &   
0.62        &    6.201986     &    6.092  &$\sim9.926^{\rm P}$ \\  
\hline
\end{tabular}
$\,$\\
\footnotesize
CM.- Mass estimated from the color--magnitud diagram.
Porb.- Photometric preiod similar to the orbital value.
M.- Age estimated from the mass--age realtionship.
P.- Age of the secondary equal to the computed age for the primary.
Msin$^3$i.- Mass computed from  M$\times$sin$^3$i.
RT.- Mass estimated from  Radii--T$_{\rm eff}$ plane. 
\end{flushleft}
\end{table*}

\newpage

\normalsize

\begin{center}
{\sc Figure Captions}
\end{center}

{Figure 1.- Color--magnitude diagram for the totality of CABS listed in 
Strassmeier et al. (1993). This figure allows the classification of these 
binaries in five different groups. Symbols for the primary and secondary
components are labeled with p and s, respectively.}

{Figure 2.- Examples of fits to the Li and Fe features.}

{Figure 3.- Lithium abundance against effective temperature. 
{\bf a} In this panel, we show our dwarf components of
CABS as solid circles, members of Hyades as open circles and 
stars belonging to Pleiades as open triangles. The average
abundance for single Hyades stars is shown as a solid line.
{\bf b}  Comparison between the abundances of Li of CABS 
--circles--, the sample of nearby  dwarfs by Favata et al. 1995 --crosses for 
actual data and triangles for upper limits-- and the lithium depletion  
isochrones by Chaboyer (1993) for 50, 70, 300, 600 and 1700 Myr 
--solid lines.}

{Figure 4.- Lithium abundance against stellar mass. {\bf a} TLBS CABS 
  are shown as solid triangles, whereas those CABS 
without the coupling are represented as solid 
circles. Those Hyades binaries with synchronizarion are shown as open squares.
The average Li abundances of Hyades stars are shown as a solid line. 
This figure eliminates the evolutionary effect on the effective temperature.
{\bf b} Comparison with NGC752.
{\bf c} Comparison with M67.
{\bf d} Comparison with NGC188.}

{Figure 5.- Li abundance against stellar age. Sizes increase with increasing 
stellar mass. Fits for single stars belonging to open clusters are shown as 
solid lines ($\sim$1.1, 1.0 and 0.9 M$\odot$). Error bars are also shown.}

{Figure 6.- Relation between Li abundances and the surface fluxes
in Ca\,{\sc ii} H\&K. Symbol sizes increase with stellar masses.}

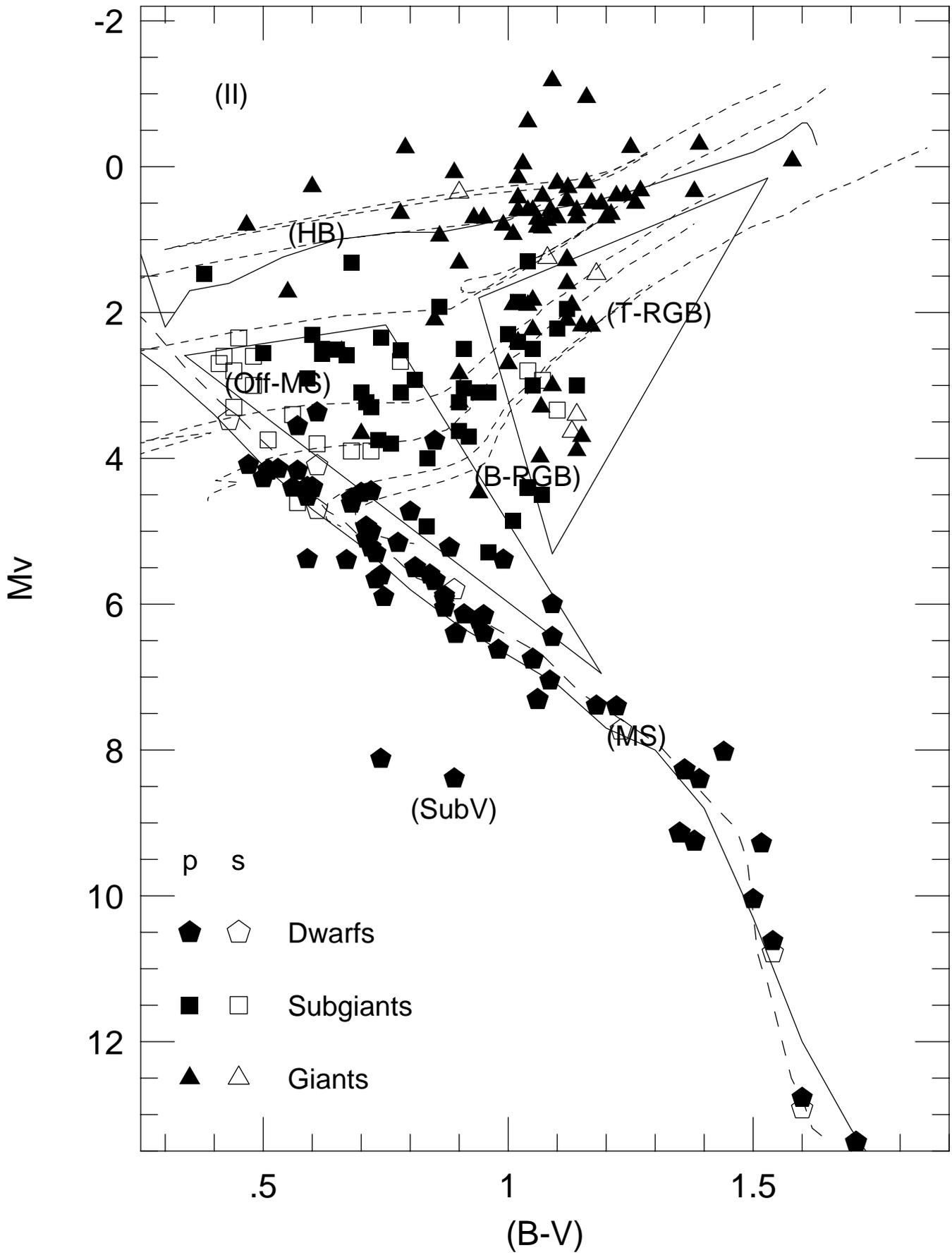
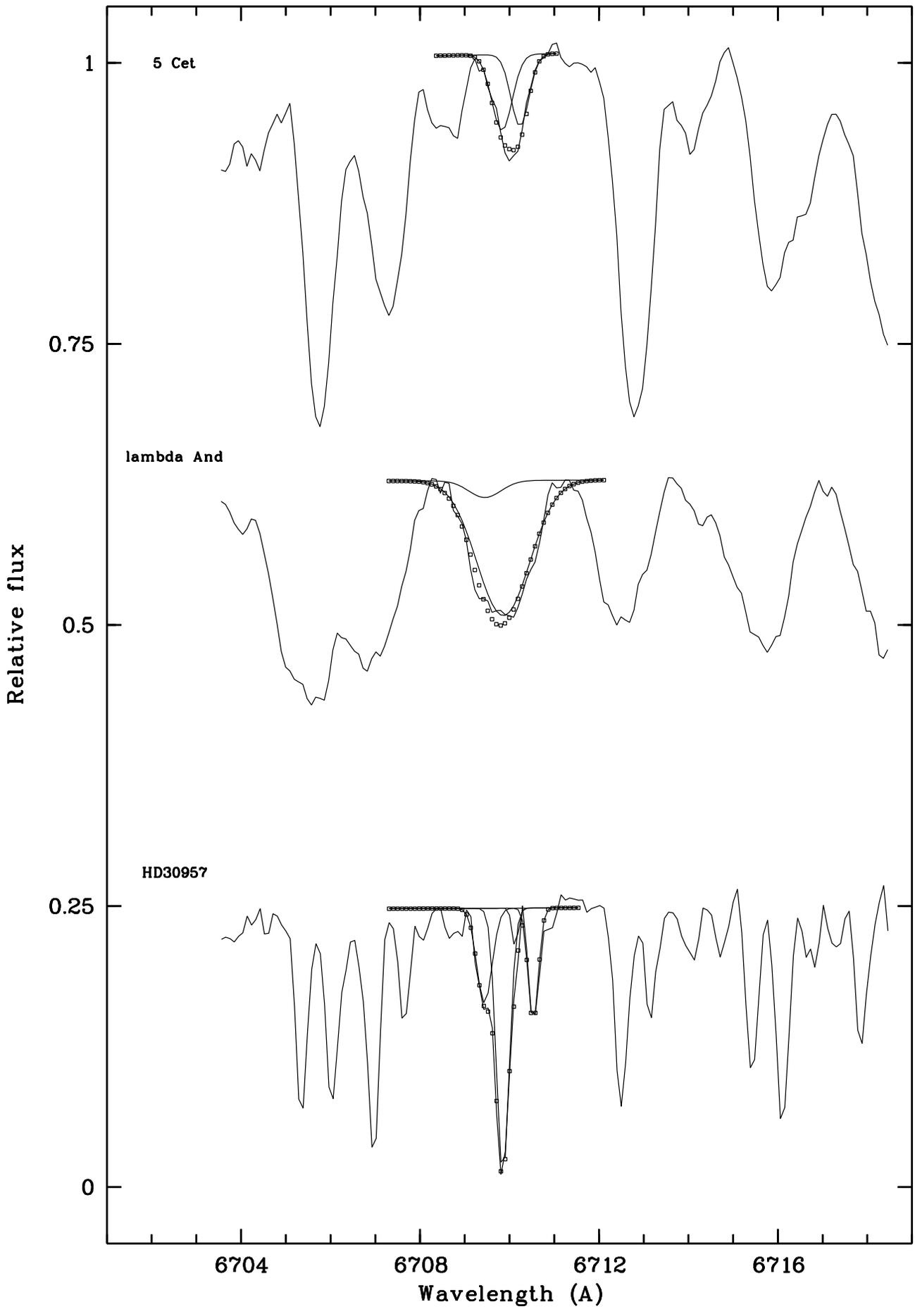
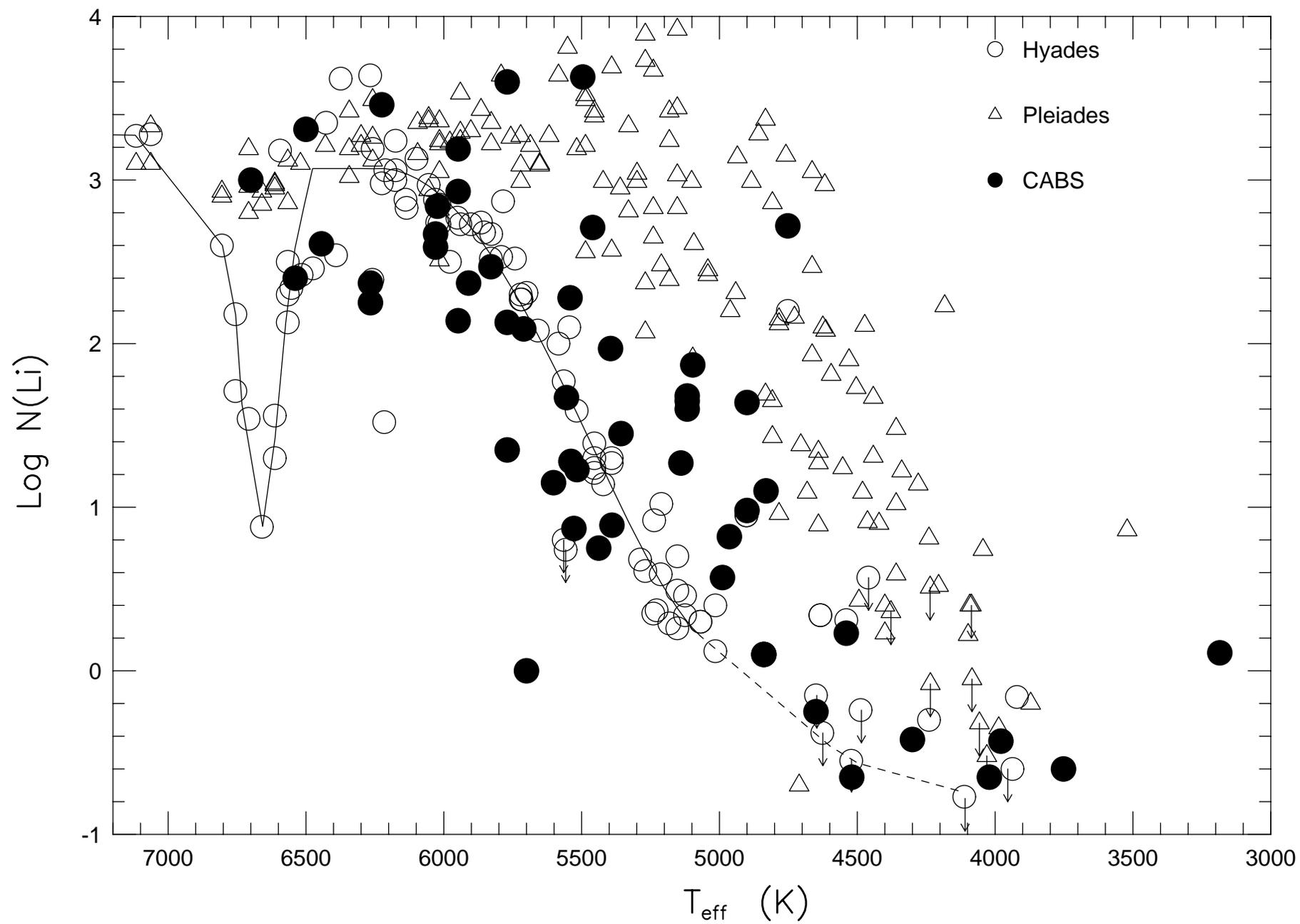
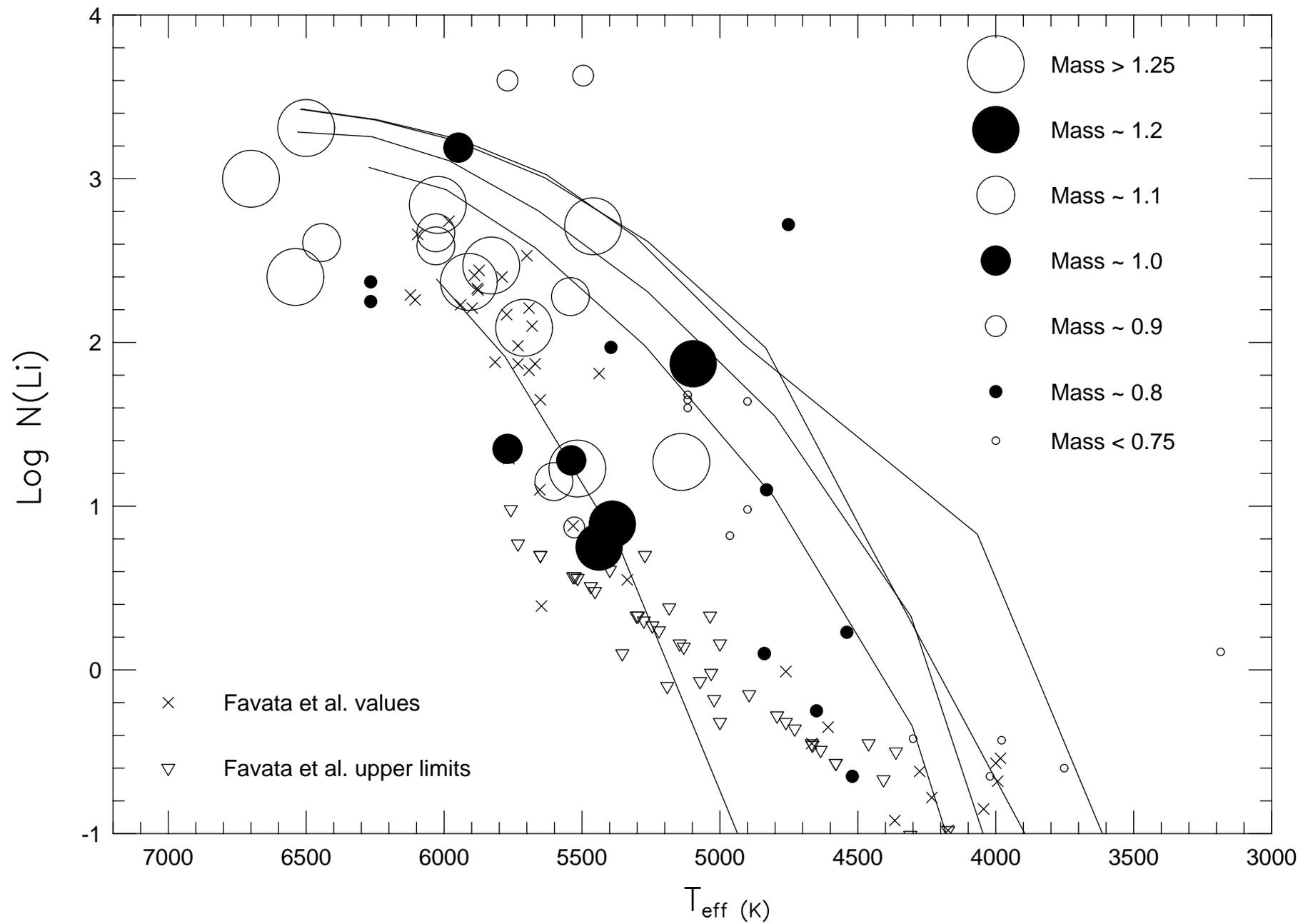
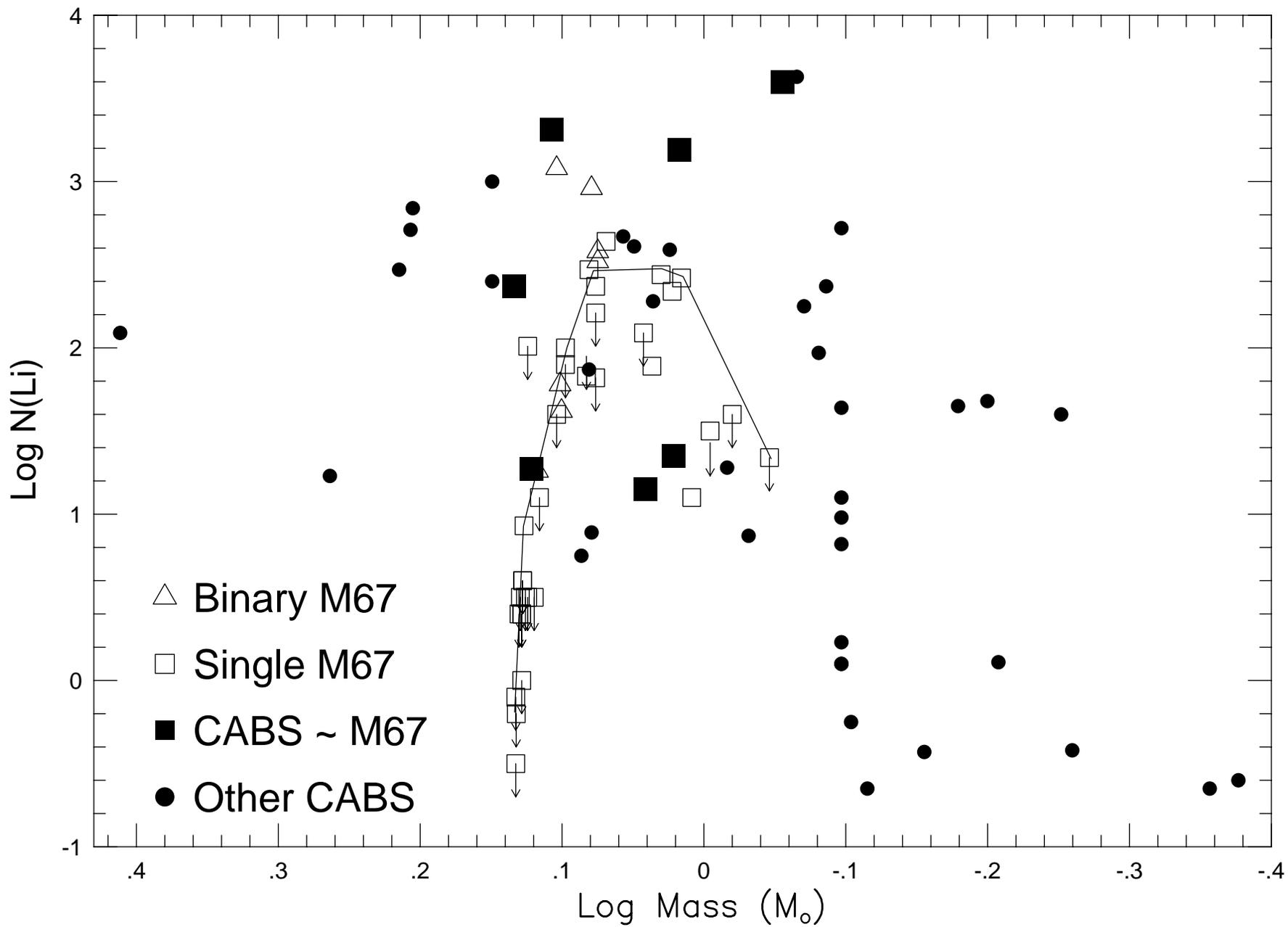
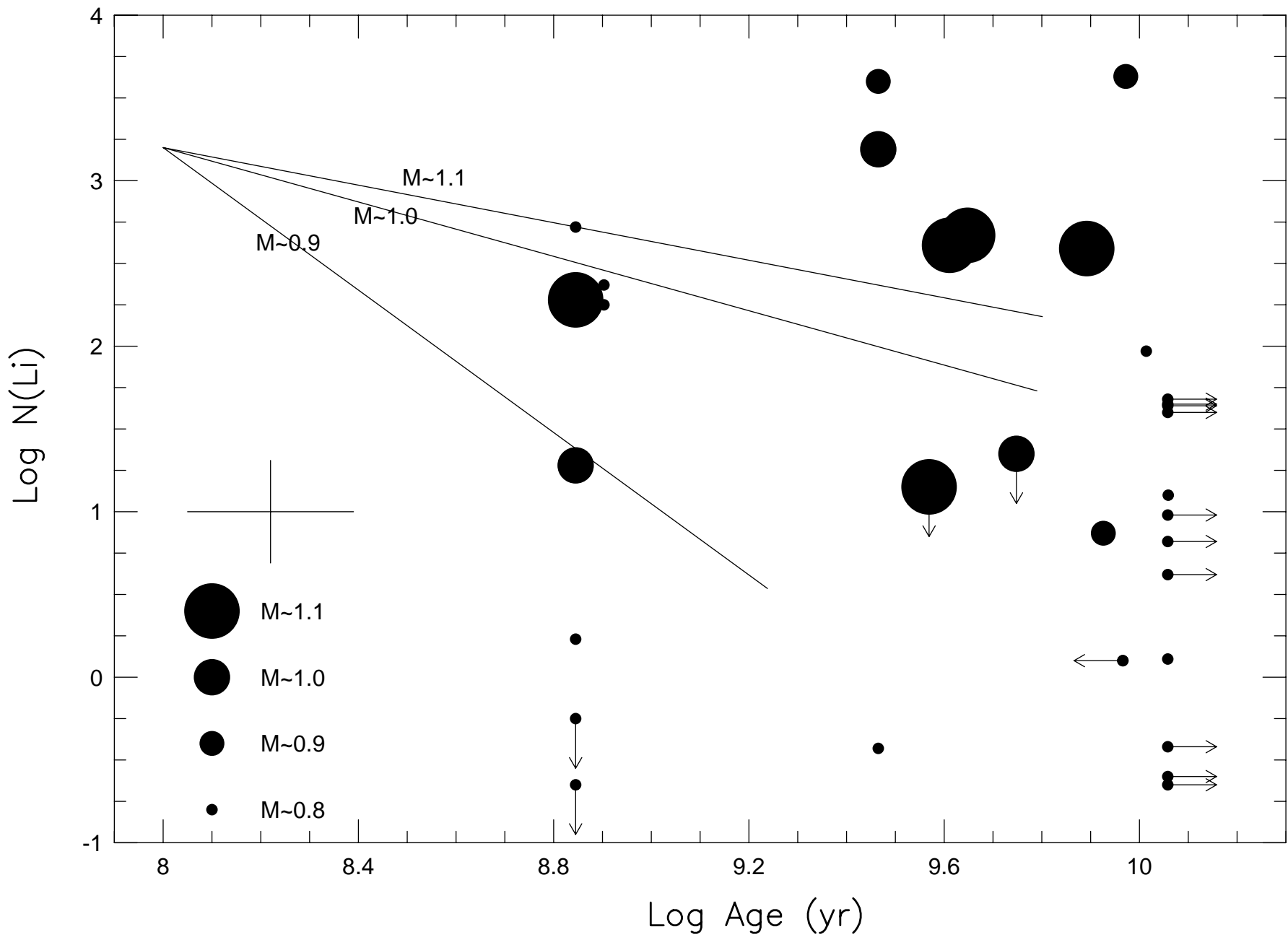
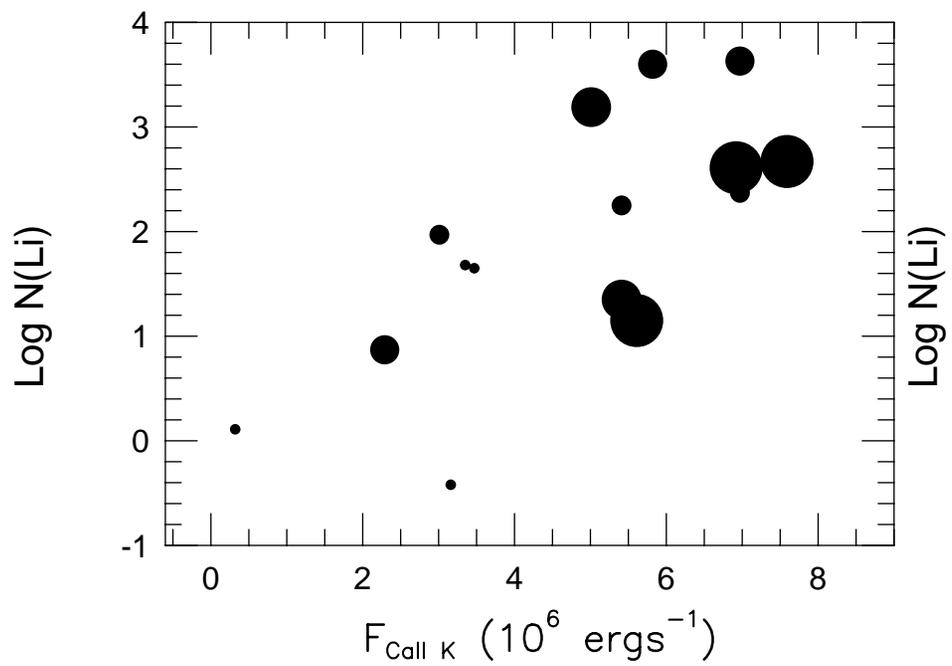
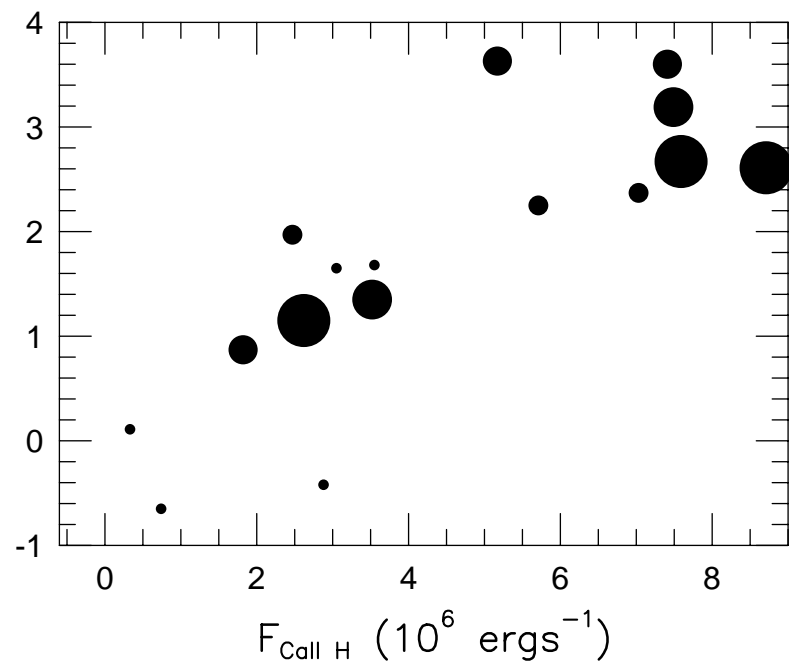
\end{document}